\documentstyle[12pt,fleqn]{article}
\begin{document}

\author{\bf{Shahin S. Agaev\thanks{Permanent address: High Energy
Physics Lab., Baku State University, Z.Khalilov st. 23, 370148 Baku,
Azerbaijan. E-mail: azhep@lan.ab.az}} \and \it{International Centre for Theoretical Physics,
Trieste, Italy}}
\title{\bf{SINGLE\ MESON\ PHOTOPRODUCTION\ AND\ IR\
RENORMALONS}}
\date{}
\maketitle

\begin{abstract}
Single pseudoscalar and vector mesons inclusive photoproduction $\gamma h\rightarrow
MX $ via higher twist mechanism is calculated using the QCD running coupling 
constant method. It is proved that in the context of this method a higher 
twist contribution to the photoproduction cross section 
cannot be normalized in terms of the meson electromagnetic form factor. 
The structure of infrared renormalon singularities of the higher twist subprocess 
cross section and the resummed expression (the Borel sum) for it 
are found. Comparisons are made with earlier results, as well as with 
leading twist cross section. Phenomenological effects of studied contributions
for $\pi,~ K,~ \rho$-meson photoproduction are discussed.
\end{abstract}
\newpage

\section{INTRODUCTION}

One of the fundamental achievements of QCD is the prediction of asymptotic
scaling laws for large-angle exclusive processes and their calculation in
the framework of perturbative QCD (pQCD) [1-3]. In the context of the
factorized QCD an expression for an amplitude of an exclusive process can be
written as integral over ${\bf x}, {\bf y}$ of hadron wave functions (w.f.)\footnote{%
Strictly speaking, $\Phi _{M}({\bf x},{\hat Q}^{2})$ is a
hadron distribution amplitude and it differs from a hadron wave function;
the former can be obtained by integrating the corresponding wave function
over partons' transverse momenta up to the factorization scale ${\hat Q}
^{2}$. But in this paper we use these two terms on the same footing.} $\Phi
_{i}({\bf x},{\hat Q}^{2})$ (an initial hadron), $\Phi_{f}^{*}({\bf x},{\hat Q}^{2})$ 
(a final hadron) and amplitude $T_{H}({\bf x},{\bf y};\alpha _{S}({\hat Q}^{2}),Q^{2})$ 
of the hard-scattering subprocess [2]. The hard-scattering amplitude 
$T_{H}({\bf x},{\bf y};\alpha _{S}({\hat Q}^{2}),Q^{2})$ depends on a process
and can be obtained in the framework of pQCD, whereas the w.f. 
$\Phi ({\bf x},{\hat Q}^{2})$ describes all the non-perturbative and 
process-independent effects of hadronic binding. The hadron w.f. gives the 
amplitude for finding partons (quarks, gluons) carrying the longitudinal 
fractional momenta ${\bf x}=(x_{1},x_{2},...x_{n})$ and virtualness up to 
${\hat Q}^{2}$ within the hadron and, in general, includes all Fock states 
with quantum numbers of the hadron. But only the lowest Fock state 
$(q_{1}\overline{q}_{2}$ - for mesons, $uud$ - for proton, etc.) contributes 
to the leading scaling behavior, other Fock states' contributions are 
suppressed by powers of $1/Q^{2}$. In our work we shall restrict ourselves by 
considering the lowest Fock state for a meson. Then, ${\bf x}=x_{1},x_{2}$
and $x_{1}+x_{2}=1.$

This approach can be applied for investigation, not only exclusive processes
but also for the calculation of higher twist (HT) corrections to some inclusive
processes, such as large-$p_{T}$ dilepton production [4], two-jet+meson
production in the electron-positron annihilation [5], etc. The HT corrections
to a single meson inclusive photoproduction and jet photoproduction 
cross sections were studied by various authors [6,7]. 
In these early papers for calculation of integrals over 
${\bf x}=x_{1},x_{2}$, like

\begin{equation}
I\sim \int \alpha _{S}(\widehat{Q}^{2})\Phi ({\bf x},\widehat{Q}%
^{2})F({\bf x}, \alpha_{S}({\hat Q}^{2}),Q^{2})\delta (1-x_{1}-x_{2})dx_{1}dx_{2}  \label{1}
\end{equation}
which appear in an expression of the amplitude, the frozen coupling constant
approximation was used. Some comments are in order concerning this point. It
is well known [8], that in pQCD calculations the argument of the running
coupling constant (or the renormalization and factorization scale) $\widehat{
Q}^{2}$ should be taken equal to the square of the momentum transfer of a
hard gluon in a corresponding Feynman diagram. But defined in this way, 
$\alpha_{S}({\hat Q}^{2})$ suffers from infrared singularities. Indeed, in a
meson form factor calculations , for example, $\widehat{Q}^{2}$ equals to $
x_{1}y_{1}Q^{2}$ or $x_{2}y_{2}Q^{2}$, $-Q^{2}$ being the four momentum square
of the virtual photon. In the single meson photoproduction 
$\gamma h\rightarrow MX$, this scale has to be chosen 
equal to $-x_{1}{\hat u}$ or $x_{2}{\hat s}$, where ${\hat u},{\hat s}$ are 
the subprocess's Mandelstam invariants [6]. Therefore, in the soft regions 
$x_{1}\rightarrow 0,y_{1}\rightarrow 0;x_{2}\rightarrow 0,y_{2}\rightarrow 0$
or $x_{1}\rightarrow 0,x_{2}\rightarrow 0$ integrals (1) diverge and for
their calculation some regularization methods of $\alpha _{S}(\widehat{Q}%
^{2}) $ in these regions are needed. In the frozen coupling approximation
these difficulties were avoided simply by equating $\widehat{Q}^{2}$ to some
fixed quantity characterizing the process. In form factor
calculations this is $\widehat{Q}^{2}\equiv Q^{2}/4$ [9], in the single meson
photoproduction -  $\widehat{Q}^{2}\equiv \widehat{s}/2,-\widehat{u}/2$ [6].

Recently, in our papers [10,11] devoted to the investigation of the light mesons
electromagnetic form factors, for their calculation we applied the running
coupling constant method, where these singularities had been
regularized by means of the principal value prescription [12]. In our recent
work we consider the inclusive photoproduction of single pseudoscalar and 
vector mesons $\gamma h\rightarrow MX$ using the same approach.

\section{CALCULATION\ OF\ THE\ HIGHER\ TWIST\ DIAGRAMS}

The two HT subprocesses, namely $\gamma q_{1}\rightarrow Mq_{2}$
and $\gamma \overline{q}_{2}\rightarrow M\overline{q}_{1}$ contribute to the
photoproduction of the single meson $M$ in the reaction $\gamma h\rightarrow
MX$ . The Feynman diagrams for the first subprocess are shown in $Fig.1$. We
do not provide the set of diagrams corresponding to the second subprocess $%
\gamma \overline{q}_{2}\rightarrow M\overline{q}_{1}$; they can be obtained
from $Fig.1$ by exchanging the quark and antiquark lines. The momenta and
charges of the particles in question are indicated in $Fig.1(a)$. In our
investigation the meson mass is neglected. As is seen from $Fig.1$, in the
HT subprocess the meson $M$ is coupled directly to the photon and
the hadron quark and its suppression in comparison with leading twist
subprocesses is caused by a hard gluon exchange in the higher twist diagrams.

The amplitude for the subprocess $\gamma q_{1}\rightarrow Mq_{2}$ can be
found by means of the Brodsky-Lepage method [2], 
\begin{equation}
M=\int_{0}^{1}\int_{0}^{1}dx_{1}dx_{2}\delta
(1-x_{1}-x_{2})T_{H}(x_{1},x_{2};\alpha _{S}(\widehat{Q}^{2}),~\widehat{s,}%
~\widehat{u},~\widehat{t})\Phi _{M}(x_{1},x_{2};\widehat{Q}^{2})  \label{2}
\end{equation}
In (2), $T_{H}$ is the sum of graphs contributing to the hard-scattering
part of the subprocess, which for the subprocess under consideration is $%
\gamma +q_{1}\rightarrow (q_{1}\overline{q}_{2})+q_{2}$, where a quark and
antiquark from the meson form a color singlet state $(q_{1}\overline{q}_{2})$%
. 

The important ingredient of our study is the choice of the meson model w.f. $%
\Phi _{M}$. In this work we calculate the photoproduction of the
pseudoscalar (pion, kaon) and vector ($\rho $-meson) mesons. For these mesons 
in the literature [3],[13] various w.f. were proposed. Here for the pion and 
kaon we use the phenomenological w.f. obtained in [3] by applying the QCD sum
rules method, for calculation of the $\rho$-meson photoproduction
we utilize both w.f. found in [3] and derived in [13]. The reason is that, in 
accordance with results of Ref.[13], the wave functions of longitudinally and
transversely polarized $\rho$-mesons are similar (coincide in shape), whereas
in [3] a significant difference between them was predicted. In [13] the authors
suggested also that the change in shape of the transverse $\rho$-meson w.f. may
increase the rate of the production of transversely polarized $\rho$-mesons by
a factor 2. We think that the single meson photoproduction is a suitable arena
for checking this conclusion.

The pion and $\rho $-meson wave functions have the form 
\begin{equation}
\Phi _{M}(x,\mu _{0}^{2})=\Phi _{asy}^{M}(x)\left[ a+b(2x-1)^{2}\right] .\
\hfill  \label{3}
\end{equation}
For the model w.f. the coefficients $a,b$ take the following values:\\
Chernyak-Zhitnitsky w.f. [3];
\begin{eqnarray}
a &=&0,b=5,~~for~ the~ pion,   \nonumber \\
a &=&0.7,b=1.5,~~for~ the~ longitudinally~ polarized~ \rho _{L}-meson,\label{4} \\
a &=&1.25,b=-1.25,~~for~ the~ transversely~ polarized~ \rho _{T}-meson.\nonumber
\end{eqnarray}
Ball-Braun w.f.[13];
$$ a=0.7,~~~b=1.5,~~$$
for both longitudinally and transversely polarized $\rho$-meson.
Here we have denoted by $x\equiv x_{1}$ the longitudinal fractional momentum
carrying by the quark within the meson. Then, $x_{2}=1-x$ and $x_{1}-x_{2}=2x-1$.

The pion and $\rho$-meson w.f. are symmetric under replacement
$x_{1}-x_{2}\leftrightarrow x_{2}-x_{1}$. But the kaon w.f. is non-symmetric;
$\Phi_{K}(x_{1}-x_{2})\neq \Phi_{K}(x_{2}-x_{1})$ [3]. Indeed, the kaon w.f. 
includes a term proportional to odd power of $(2x-1)$, 
\begin{eqnarray}
\Phi _{K}(x,\mu _{0}^{2}) &=&\Phi _{asy}^{K}(x)\left[
a+b(2x-1)^{2}+c(2x-1)^{3}\right] ,  \label{5} \\
a &=&0.4,~~b=3,~~c=1.25,  \nonumber
\end{eqnarray}
and may be written as the sum of the symmetric $\Phi _{s}(x,\mu _{0}^{2})$
and antisymmetric $\Phi _{a}(x,\mu _{0}^{2})$ parts,

\begin{equation}
\Phi _{s}(x,\mu _{0}^{2})=\Phi _{asy}^{K}(x)\left[ a+b(2x-1)^{2}\right]
,~~~\Phi _{a}(x,\mu _{0}^{2})=\Phi _{asy}^{K}(x)c(2x-1)^{3}. \label{6}
\end{equation}
In (3),(5),(6) $\Phi _{asy}^{M}(x)$ is the asymptotic w.f. 
\begin{equation}
\Phi _{asy}^{M}(x)=\sqrt{3}f_{M}x(1-x),  \label{7}
\end{equation}
where $f_{M}$ is the meson decay constant; $f_{\pi
}=0.093~GeV,f_{K}=0.112~GeV$. In the case of the $\rho$-meson we take
$f^{L}_{\rho }=f^{T}_{\rho }=0.2~GeV$ for the CZ w.f., and $f_{\rho}^{L}=0.2~GeV$, 
$f_{\rho}^{T}=0.16~GeV$ for BB w.f.

The normalization of $\Phi _{M}(x,\mu _{0}^{2})$ at $\mu _{0}=0.5~GeV$ is
given by the condition 
\begin{equation}
\int_{0}^{1}dx\Phi _{M}(x,\mu _{0}^{2})=\frac{f_{M}}{2\sqrt{3}}.  \label{8}
\end{equation}
The factor $\sqrt{2}$ appearing in the normalization of a vector meson is 
included in the $\rho$-meson decay constant.

The formalism for calculation of the HT subprocess cross section
is well known and described in [6,14]. We omit details of our
calculations and write down the final expression for $d\widehat{\sigma }%
^{HT}/d\widehat{t}$. We find:

for the pseudoscalar and longitudinally polarized vector mesons,

\begin{eqnarray}
\frac{d\widehat{\sigma }^{HT}(e_{1},e_{2})}{d\widehat{t}} &=&\frac{32\pi
^{2}C_{F}\alpha _{E}}{9\widehat{s}^{2}}\left\{ -\frac{e_{1}^{2}}{\widehat{s}%
^{2}}\left[ I_{1}^{2}\widehat{t}-2I_{1}\left( I_{1}\widehat{s}+I_{2}\widehat{%
u}\right) \frac{\widehat{u}}{\widehat{t}}+I_{2}^{2}\frac{\widehat{u}^{2}}{%
\widehat{t}}\right] -\right.  \nonumber \\
&&\frac{e_{2}^{2}}{\widehat{u}^{2}}\left[ K_{1}^{2}\widehat{t}-2K_{1}\left(
K_{1}\widehat{u}+K_{2}\widehat{s}\right) \frac{\widehat{s}}{\widehat{t}}%
+K_{2}^{2}\frac{\widehat{s}^{2}}{\widehat{t}}\right] -  \label{9} \\
&&\left. \frac{2e_{1}e_{2}}{\widehat{s}\widehat{u}\widehat{t}}\left[
I_{1}K_{1}\widehat{t}^{2}-I_{1}(K_{2}\widehat{s}+K_{1}\widehat{u})\widehat{s}%
-K_{1}(I_{1}\widehat{s}+I_{2}\widehat{u})\widehat{u}\right] \right\} . 
\nonumber
\end{eqnarray}

for the transversely polarized vector meson,

\begin{equation}
\frac{d{\hat \sigma}^{HT}(e_{1},e_{2})}{d{\hat t}}=
\frac{64\pi^{2}C_{F}\alpha_{E}}{9{\hat s}^{4}}\frac{-{\hat t}}{{\hat u}^{2}}
\left [e_{1}{\hat u}I_{2}-e_{2}{\hat s}K_{2}\right ]^{2}
\end{equation}

In (9),(10), $\alpha _{E}\simeq 1/137$ is the fine structure constant, $%
C_{F}=4/3$ is the color factor. The Mandelstam invariants for the subprocess
are defined as 
\begin{eqnarray}
\widehat{s} &=&(zp+q)^{2}=zs,  \nonumber \\
\widehat{t} &=&(q-P)^{2}=t,  \label{11} \\
\widehat{u} &=&(zp-P)^{2}=zu,  \nonumber
\end{eqnarray}
where $s,t,u$ are the Mandelstam invariants for the process $\gamma
h\rightarrow MX$ , $z$ is the longitudinal fractional momentum of the quark $%
q_{1}$ out of the hadron $h$.

The main problem in our investigation is the calculation of quantities $I_{1,2}$%
, $K_{1,2}$,

\begin{equation}
I_{1}=\int_{0}^{1}\int_{0}^{1}\frac{dx_{1}dx_{2}\delta (1-x_{1}-x_{2})\alpha
_{S}(\widehat{Q}_{1}^{2})\Phi _{M}(x_{1},x_{2};\widehat{Q}_{1}^{2})}{x_{2}},
\label{12}
\end{equation}

\begin{equation}
I_{2}=\int_{0}^{1}\int_{0}^{1}\frac{dx_{1}dx_{2}\delta (1-x_{1}-x_{2})\alpha
_{S}(\widehat{Q}_{1}^{2})\Phi _{M}(x_{1},x_{2};\widehat{Q}_{1}^{2})}{%
x_{1}x_{2}},  \label{13}
\end{equation}
and 
\begin{equation}
K_{1}=\int_{0}^{1}\int_{0}^{1}\frac{dx_{1}dx_{2}\delta (1-x_{1}-x_{2})\alpha
_{S}(\widehat{Q}_{2}^{2})\Phi _{M}(x_{1},x_{2};\widehat{Q}_{2}^{2})}{x_{1}},
\label{14}
\end{equation}

\begin{equation}
K_{2}=\int_{0}^{1}\int_{0}^{1}\frac{dx_{1}dx_{2}\delta (1-x_{1}-x_{2})\alpha
_{S}(\widehat{Q}_{2}^{2})\Phi _{M}(x_{1},x_{2};\widehat{Q}_{2}^{2})}{x_{2}},
\label{15}
\end{equation}
where for $I_{1},I_{2}$ the renormalization and factorization scale is $%
\widehat{Q}_{1}^{2}=x_{2}\widehat{s}$, for $K_{1},K_{2}$ it is given by $%
\widehat{Q}_{2}^{2}=-x_{1}\widehat{u}$.

Let us first consider the frozen coupling constant approximation. In this
approximation we put $\widehat{Q}_{1,2}^{2}$ equal to their mean values $%
\widehat{s}/2,-\widehat{u}/2$ and remove $\alpha _{S}(\widehat{Q}_{1,2}^{2})$
as the constant factor in (12-15). After such manipulation, the integrals
(12-15) are trivial and can easily be found. For the mesons with symmetric
w.f. we get 
\begin{eqnarray*}
I_{2}^{0}\left( \frac{\widehat{s}}{2}\right)  &=&2I_{1}^{0}\left( \frac{%
\widehat{s}}{2}\right) \equiv \alpha _{S}\left( \frac{\widehat{s}}{2}\right)
I_{M}\left( \frac{\widehat{s}}{2}\right) ,\hspace{0.5in} \\
K_{2}^{0}\left( -\frac{\widehat{u}}{2}\right)  &=&2K_{1}^{0}\left( -\frac{%
\widehat{u}}{2}\right) \equiv \alpha _{S}\left( -\frac{\widehat{u}}{2}%
\right) I_{M}\left( -\frac{\widehat{u}}{2}\right) ,
\end{eqnarray*}
where superscript ''$0$'' indicates that the quantities $I,K$ are found in
the frozen coupling approximation. Here the function $I_{M}(\widehat{Q}^{2})$
is 
$$I_{M}(\widehat{Q}^{2})=\int_{0}^{1}\frac{dx\Phi _{M}(x,\widehat{Q}^{2})}{%
x(1-x)}.$$
In this approximation, using the last expressions and (9),(10),(12-15)
one can easily reproduce results of [6] for the subprocess cross section$%
\footnote{%
The difference between our expressions and corresponding formulas in Ref.[6]
is caused by our definition of the antiquark's charge, i.e. in our
expressions the charge of the antiquark from $M$ is $-e_{2}$, whereas in
Ref.[6] it is denoted by $e_{2}$.}$.

In the case of the kaon we find

\begin{eqnarray}
I_{1}^{0}\left( \frac{\widehat{s}}{2}\right)  &=&\alpha _{S}\left( \frac{%
\widehat{s}}{2}\right) \left[ \int_{0}^{1}\frac{dx\Phi _{s}(x,\widehat{s}/2)%
}{1-x}+\int_{0}^{1}\frac{dx\Phi _{a}(x,\widehat{s}/2)}{1-x}\right] ,
\label{16} \\
I_{2}^{0}\left( \frac{\widehat{s}}{2}\right)  &=&2\alpha _{S}\left( \frac{%
\widehat{s}}{2}\right) \int_{0}^{1}\frac{dx\Phi _{s}(x,\widehat{s}/2)}{1-x}.
\nonumber
\end{eqnarray}
It is evident that

$$I_{2}^{0}(\widehat{s}/2)\neq 2I_{1}^{0}(\widehat{s}/2).$$

The same is also true for $K_{1}^{0}$ and $K_{2}^{0}$. This means that in
the case of a pseudoscalar meson with the non-symmetric w.f. the result of
[6] is not valid and in the calculations our expressions (9),(12-15) have to 
be applied.

The important problem in the single meson inclusive photoproduction is the possibility of
normalization of the HT subprocess cross section (9),(10) in
terms of the electromagnetic form factor $F_{M}(Q^{2})$ of the corresponding
meson.

The electromagnetic form factor $F_{M}(Q^{2})$ of the meson $M$ is given by
the expression 
\begin{equation}
F_{M}(Q^{2})=\int_{0}^{1}\int_{0}^{1}\Phi _{M}^{*}(y,\widehat{Q}%
^{2})T_{H}^{ff}(x,y;\alpha _{S}(\widehat{Q}^{2}),Q^{2})\Phi _{M}(x,\widehat{Q%
}^{2})dxdy.  \label{17}
\end{equation}
Here

\[
T_{H}^{ff}(x,y;\alpha _{S}(\widehat{Q}^{2}),Q^{2})=\frac{16\pi C_{F}}{Q^{2}}%
\left[ e_{1}\frac{\alpha _{S}(Q^{2}(1-x)(1-y))}{(1-x)(1-y)}-e_{2}\frac{%
\alpha _{S}(Q^{2}xy)}{xy}\right] . 
\]
For the meson with symmetric w.f. using the frozen coupling approximation ($%
\widehat{Q}^{2}\rightarrow Q^{2}/4$) we get

\begin{equation}
F_{M}(Q^{2})=\frac{16\pi C_{F}\alpha _{S}(Q^{2}/4)}{Q^{2}}%
(e_{1}-e_{2})\left( \int_{0}^{1}\frac{\Phi _{M}(x,Q^{2}/4)dx}{1-x}\right)
^{2}.  \label{18}
\end{equation}
It is not difficult to conclude that for such mesons the
quantities $(I^{0})^{2}$ and $(K^{0})^{2}$ in the cross section can
be expressed in terms of $F_{M}$%
\begin{equation}
\left[ I_{1}^{0}(\widehat{s}/2)\right] ^{2}=\left. \frac{\alpha _{S}(%
\widehat{s}/2)}{16\pi C_{F}}\left[ Q^{2}\left| F_{M}(Q^{2})\right| \right]
\right| _{{\hat Q}^{2}=2{\hat s}}
\end{equation}

\[
\left[ I_{2}^{0}(\widehat{s}/2)\right] ^{2}=\left. \frac{\alpha _{S}(%
\widehat{s}/2)}{4\pi C_{F}}\left[ Q^{2}\left| F_{M}(Q^{2})\right| \right]
\right| _{{\hat Q}^{2}=2{\hat s}}
\]
For mesons with non-symmetric w.f. from (17) we find

\[
F_{M}(Q^{2})=\frac{16\pi C_{F}\alpha _{S}(Q^{2}/4)}{Q^{2}}\left\{
(e_{1}-e_{2})\left( \int_{0}^{1}\frac{\Phi _{s}(x,Q^{2}/4)dx}{1-x}\right)
^{2}+\right. 
\]

\[
(e_{1}-e_{2})\left( \int_{0}^{1}\frac{\Phi _{a}(x,Q^{2}/4)dx}{1-x}\right)
^{2}+ 
\]

\begin{equation}
\left. 2(e_{1}+e_{2})\left( \int_{0}^{1}\frac{\Phi _{s}(x,Q^{2}/4)dx}{1-x}%
\right) \left( \int_{0}^{1}\frac{\Phi _{a}(x,Q^{2}/4)dx}{1-x}\right)
\right\} .  \label{20}
\end{equation}
It is now clear that $(I^{0})^{2},(K^{0})^{2}$ (16) are not
proportional to $F_{M}$ (20). This means that even in the context of
the frozen coupling approximation the HT subprocess cross section
may be normalized in terms of the meson form factor only if the photoproduction
of the meson with symmetric w.f. is considered.

\section{THE\ RUNNING\ COUPLING\ CONSTANT\ METHOD\ AND\ IR\ RENORMALONS}

In this section we shall calculate the integrals (12-15) using the running
coupling constant method and also discuss the problem of
normalization of the higher twist process cross section in terms of the
meson electromagnetic form factor obtained in the context of the same approach.

As is seen from (12-15), in general, one has to take into account not
only the dependence of $\alpha (\widehat{Q}_{1,2}^{2})$ on the scale $\widehat{Q}_{1,2}^{2}$, but also an evolution of $%
\Phi _{M}(x,\widehat{Q}_{1,2}^{2})$ with $\widehat{Q}_{1,2}^{2}$. The meson
w.f. evolves in accordance with a Bethe-Salpeter type equation, but its
dependence on $\widehat{Q}^{2}$ is mild and may be neglected by replacing 
$\Phi _{M}(x,\widehat{Q}_{1,2}^{2})\rightarrow \Phi _{M}(x,\mu _{0}^{2})$.
Such approximation does not change considerably numerical results, but
phenomenon considering in this article (effect of infrared renormalons)
becomes transparent.

Let us clarify our method by calculating the integral (12); the quantities $%
I_{2},K_{1,2}$ can be worked out in the same way. For the mesons with
symmetric w.f. Eq.(12) in the framework of the running coupling approach
takes the form

\begin{equation}
I_{1}(\widehat{s})=\int_{0}^{1}\frac{\alpha _{S}((1-x)\widehat{s})\Phi
_{M}(x,\mu _{0}^{2})dx}{1-x}.  \label{21}
\end{equation}
The $\alpha _{S}((1-x)\widehat{s})$ has the infrared singularity at $%
x\rightarrow 1$ and as a result integral (21) diverges (the pole associated
with the denominator of the integrand is fictitious, because $\Phi _{M}\sim
(1-x),$ and therefore, the singularity of the integrand at $x=1$ is caused
only by $\alpha _{S}((1-x)\widehat{s})$). For the regularization of the integral
let us relate the running coupling at scaling variable 
$\alpha_{S}((1-x){\hat s})$ with the aid of the renormalization group equation 
in terms of the fixed one $\alpha _{S}(\widehat{s})$. The renormalization group equation for
the running coupling $\alpha (\widehat{s})\equiv $ $\alpha _{S}(\widehat{s}%
)/\pi $ 

\begin{equation}
\frac{\partial \alpha (\lambda \widehat{s})}{\partial \ln \lambda }\simeq -%
\frac{\beta _{0}}{4}\left[ \alpha (\lambda \widehat{s})\right] ^{2},
\label{22}
\end{equation}
has the solution [12] 
\begin{equation}
\alpha(\lambda \widehat{s})\simeq \frac{\alpha(\widehat{s})}{%
1+(\alpha (\widehat{s})\beta _{0}/4)\ln \lambda }.  \label{23}
\end{equation}
In (22),(23), the one-loop QCD coupling constant $\alpha_{S}(\mu^{2})$ is
defined as
$$\alpha_{S}(\mu^{2})=\frac{4\pi}{\beta_{0}\ln(\mu^{2}/\Lambda^{2})}$$
$\beta_{0}=11-2n_{f}/3$ being the QCD beta-function first coefficient.

Having inserted (23) into (21) we get 
\begin{equation}
I_{1}(\widehat{s})=\alpha _{S}(\widehat{s})\int_{0}^{1}\frac{\Phi _{M}(x,\mu
_{0}^{2})dx}{(1-x)(1+(1/t)\ln (1-x))},  \label{24}
\end{equation}
where $t=4\pi /\alpha _{S}(\widehat{s})\beta _{0}$.\\
The integral (24) is, of course, still divergent, but now it is recasted into
a form, which is suitable for calculation. Using the method described in
details in our work [10] it may be found as a perturbative series in 
$\alpha _{S}(\widehat{s})$

\begin{equation}
I_{1}(\widehat{s})\sim \sum_{n=1}^{\infty}\left( 
\frac{\alpha _{S}(\widehat{s})}{4\pi }\right) ^{n}S_{n},~~~S_{n}=C_{n}\beta_{0}^{n-1}.
\label{25}
\end{equation}
The coefficients $C_{n}$ of this series demonstrate factorial growth $C_{n}\sim (n-1)!$, which
might indicate an infrared renormalon nature of divergences in the integral
(24) and corresponding series (25). The procedure for dealing with such
ill-defined series is well known; one has to perform the Borel transform of
the series [15]

\begin{equation}
B[I_{1}](u)=\sum_{n=1}^{\infty}\frac{u^{n-1}}{(n-1)!}%
C_{n},  \label{26}
\end{equation}
then invert $B[I_{1}](u)$ to obtain the resummed expression (the
Borel sum) for $I_{1}(\widehat{s})$. This method is straightforward but
tedious. Therefore, it is convenient to apply the second method proposed
also in our work [11], which allows us to bypass all these intermediate
steps and find directly the resummed expression for $I_{1}(\widehat{s})$.
For these purposes let us introduce the inverse Laplace transform of $%
1/(t+z)$

\begin{equation}
\frac{1}{t+z}=\int_{0}^{\infty}\exp [-(t+z)u]du.
\label{27}
\end{equation}
Then $I_{1}({\hat s})$ may be readily carried out by the change of the variable $x$ to 
$z=\ln (1-x)$ and using (27)

\begin{eqnarray}
I_{1}(\widehat{s})&=&\frac{4\sqrt{3}\pi f_{M}}{\beta _{0}}
\int_{0}^{\infty}\exp \left[ -\frac{4\pi u}{\alpha _{S}(\widehat{s})
\beta _{0}}\right] \left( \frac{a+b}{1-u}-\frac{a+5b}{2-u}+ \right.
\nonumber  \\
&&\left. \frac{8b}{3-u}-\frac{4b}{4-u}\right ) .\label{28}
\end{eqnarray}
Eq.(28) is nothing more than the Borel sum of
the perturbative series (25) and the corresponding Borel transform is 

\begin{equation}
B[I_{1}](u)=\frac{a+b}{1-u}-\frac{a+5b}{2-u}+\frac{8b}{3-u}-\frac{4b}{4-u}.
\label{29}
\end{equation}
The series (25) can be recovered by means of the following formula
$$C_{n}=\left. \left( \frac{d}{du}\right) ^{n-1}B[I_{1}](u)\right| _{u=0}.$$

The Borel transform $B[I_{1}](u)$ has poles on the real $u$ axis at $%
u=1;2;3;4$, which confirms our conclusion concerning the infrared
renormalon nature of divergences in (25). To remove them from Eq.(28)
some regularization methods have to be applied. In this article we adopt the
principal value prescription [12]. We obtain

\begin{eqnarray}
\left[ I_{1}\left( \widehat{s}\right) \right] ^{res}&=&\frac{4\sqrt{3}\pi f_{M}%
}{\beta _{0}}\left[ \left( a+b\right) \frac{Li(\lambda )}{\lambda }-\left(
a+5b\right) \frac{Li(\lambda ^{2})}{\lambda ^{2}}+\right.
\nonumber\\
&&\left. 8b\frac{Li(\lambda ^{3})}{%
\lambda ^{3}}-4b\frac{Li(\lambda ^{4})}{\lambda ^{4}}\right] ,  \label{30}
\end{eqnarray}
where $Li(\lambda )$ is the logarithmic integral [16], for $\lambda >1$
defined in its principal value
\begin{equation}
Li(\lambda )=P.V.\int_{0}^{\infty}\frac{dx}{\ln x}%
,~~~~\lambda =\widehat{s}/\Lambda ^{2}.  \label{31}
\end{equation}
For other integrals from (13-15) we find
\begin{equation}
\left[ I_{2}\left( \widehat{s}\right) \right] ^{res}=\frac{4\sqrt{3}\pi f_{M}%
}{\beta _{0}}\left[ \left( a+b\right) \frac{Li(\lambda )}{\lambda }-4b\frac{%
Li(\lambda ^{2})}{\lambda ^{2}}+4b\frac{Li(\lambda ^{3})}{\lambda ^{3}}%
\right] ,  \label{32}
\end{equation}
and
\begin{equation}
\left[ K_{1}\left( -\widehat{u}\right) \right] ^{res}=\left[ I_{1}\left( -%
\widehat{u}\right) \right] ^{res},\left[ K_{2}\left( -\widehat{u}\right)
\right] ^{res}=\left[ I_{2}\left( -\widehat{u}\right) \right] ^{res}.
\label{33}
\end{equation}
From (30),(32),(33), we conclude that in the framework of the running
coupling approximation even for mesons with symmetric w.f. we have 
$$\left[ I_{2}\left( \widehat{s}\right) \right] ^{res}\not\sim \left[ I_{1}\left( 
\widehat{s}\right) \right] ^{res},~~~\left[ K_{2}\left( -\widehat{u}\right)
\right] ^{res}\not\sim \left[ K_{1}\left( -\widehat{u}\right) \right] ^{res}.$$
Therefore only our results for the subprocess cross section (9),(10)
are correct.

Another question is, as we have discussed in Sect.2, the normalization of the
meson photoproduction cross section in terms of the meson elm form
factor. The pion and kaon form factors have been calculated by means of the
running coupling approach in our previous papers [10, 11]. Let us write down
the pion form factor obtained using the pion's simplest w.f., that is, the
asymptotic one ($a=1,b=0$ in (4))

\begin{equation}
\left[ Q^{2}F_{\pi }(Q^{2})\right] _{asy}^{res}=\frac{(16\pi f_{\pi })^{2}}{%
\beta _{0}}\left[ -\frac{3}{2}+(\ln \lambda -2)\frac{Li(\lambda )}{\lambda }%
+(\ln \lambda +2)\frac{Li(\lambda ^{2})}{\lambda ^{2}}\right] .  \label{34}
\end{equation}
From (30),(34) it follows that the relations (19) do no longer hold. The same
is also true for the pion's other w.f., as well as for $\rho _{L}-$ and $%
\rho _{T}-$mesons. In other words, in the running coupling approach the
HT subprocess cross section (9),(10) cannot be normalized in
terms of the meson form factor neither for mesons with symmetric w.f. nor
for non-symmetric ones.

Let us, for completeness, write down $I({\hat s}),K(-{\hat u})$ calculated for 
non-symmetric w.f. (5)
\begin{eqnarray}
\left[ I_{1}\left( \widehat{s}\right) \right] ^{res}&=&\frac{4\sqrt{3}\pi f_{M}%
}{\beta _{0}}\left[ \left( a+b+c\right) \frac{Li(\lambda )}{\lambda }-\left(
a+5b+7c\right) \frac{Li(\lambda ^{2})}{\lambda ^{2}}+\right. 
\nonumber\\
&&\left. 2(4b+9c)\frac{Li(\lambda ^{3})}{\lambda ^{3}}-4(b+5c)\frac{Li(\lambda
^{4})}{\lambda ^{4}}+8c\frac{Li(\lambda ^{5})}{\lambda ^{5}}\right] ,
\label{35}
\end{eqnarray}
\begin{eqnarray}
\left[ I_{1}\left( \widehat{s}\right) \right] ^{res} &=&\frac{4\sqrt{3}\pi
f_{M}}{\beta _{0}}\left[ \left( a+b+c\right) \frac{Li(\lambda )}{\lambda }%
-2\left( 2b+3c\right) \frac{Li(\lambda ^{2})}{\lambda ^{2}}+\right.  
\nonumber \\
&&\left. 4(b+3c)\frac{Li(\lambda ^{3})}{\lambda ^{3}}-8c\frac{Li(\lambda
^{4})}{\lambda ^{4}}\right] .  \label{36}
\end{eqnarray}
The expressions for $\left[ K_{1}\left( -\widehat{u}\right) \right] ^{res}$
and $\left[ K_{2}\left( -\widehat{u}\right) \right] ^{res}$ may be obtained
from (35),(36) by $c\rightarrow -c,\lambda =\widehat{s}/\Lambda
^{2}\rightarrow -\widehat{u}/\Lambda ^{2}$ replacements, respectively. With
these explicit expressions and the results of [11] at hand one can check
our statements concerning the normalization of the subprocess cross section 
for kaons.

Some comments are in order concerning these results. First of all, it is 
instructive to compare sources of the infrared renormalons in our case and 
in other QCD processes considered in [17]. In these articles the running coupling
constant method was used in one-loop order calculations for resummation of any 
number of fermion bubble insertions in the gluon propagator. This technique
corresponds to partial resummation of the perturbative series for a quantity
under consideration. Indeed, for such quantities, the coefficients $S_{n}$ in 
(25) have the following expansion in powers of $\beta_{0}$,
$$S_{n}=C_{n}\beta_{0}^{n-1}+R_{n}\beta_{0}^{n-2}+\ldots~.$$
Therefore, by defining the Borel transform as in (26) and inverting it one 
obtains a partially resummed expression for a physical quantity. In our 
calculations of the HT cross section we use the leading order term for $T_{H}$;
there are no gluon loops in the corresponding Feynman diagrams in $Fig.1$. 
It is not accidental that $S_{n}$ (25) in our case has exactly $\sim
\beta_{0}^{n-1}$ dependence and, hence, the expressions (30),(32),(35),(36)
are exact sums of the corresponding perturbative series.
In [17] it has been demonstrated that the only source of terms of order 
$\sim \alpha_{S}^{n}\beta_{0}^{n-1}$ in (25) is the running coupling constant.
The source of these terms in the HT calculations is also the running coupling
$\alpha_{S}({\hat s}(1-x))$ (or $\alpha_{S}(-{\hat u}(1-x))$), which runs due
to the integration in (1) over the meson quark's (antiquark's) longitudinal
fractional momentum, but not because of a loop integration.

Another question commonly discussed in papers involving IR renormalons is an
ambiguity produced by the principal value prescription used for the regularization
of divergent integrals (28). The ambiguity introduced by our treatment of (28)
is a higher twist and behaves as $\Lambda^{2}/{\hat Q}^{2}$ (the first renormalon
pole is $u=1$). But the subprocess under consideration itself is already the
higher twist one. Therefore, we can safely ignore such "HT-to-HT" corrections.

At the end of this section let us write down the HT correction to the
single meson photoproduction
cross section by taking into account both HT subprocesses; $\gamma 
q_{1}\rightarrow Mq_{2}$ and $\gamma {\bar q}_{2}\rightarrow M{\bar q}_{1}$. It
is not difficult to prove that the second subprocess cross section can be
obtained from (9),(10) by $e_{1}\leftrightarrow e_{2}$ replacement. Then
the HT correction to the single meson photoproduction cross section is given by
\begin{eqnarray}
\frac{\sigma^{HT}}{dp_{T}^{2}dy}&=&z^{*}\sum_{q_{1}, {\bar q}_{2}}
\left\{ q_{1}^{h}(z^{*},-t) \frac{d {\hat \sigma}^{HT}(e_{1},e_{2})}
{d{\hat t}}+\right. 
\nonumber \\
&& \left. {\bar q}_{2}^{h}(z^{*},-t) \frac{d {\hat \sigma}^{HT}(e_{2},e_{1})}
{d {\hat t}}\right\}\frac{s}{s+u}. \label{50} 
\end{eqnarray}
where
$$z^{*}=\frac{p_{T}e^{-y}}{\sqrt{s}-p_{T}e^{y}}.$$
Here the sum runs over the hadron's quark $q_{1}$ and antiquark $\overline{q}%
_{2}$ flavors. In (37) $q_{1}^{h}(z^{*},-t)$, $\overline{q}%
_{2}^{h}(z^{*},-t)$ are the quark and antiquark distribution functions,
respectively. All r.h.s. quantities are expressed in terms of the process
c.m. energy $\sqrt{s}$, the meson transverse momentum $p_{T}$ and rapidity $%
y $ using the following expressions

\begin{equation}
{\hat s}=\frac{sp_{T}e^{-y}}{\sqrt{s}-p_{T}e^{y}},~~
{\hat t}=-p_{T}\sqrt{s}e^{-y},~~{\hat u}=-\frac{p_{T}^{2}\sqrt{s}}{\sqrt{s}-p_{T}e^{y}}.
\end{equation}
Eq.(37) is the final result which will be used later in our numerical
calculations.

\section{PHOTOPRODUCTION\ OF\ MESONS\ AT\ THE\ LEADING\ TWIST\ LEVEL}

In our study of the single meson photoproduction a crucial point is
the comparison of our results with leading twist (LT) ones. This will enable us 
to find such domains in the phase space in which the higher twist
photoproduction mechanism is actually observable.

The LT subprocesses, which contribute to a meson photoproduction
are:\\ 
a photon-quark (antiquark) scattering
\begin{equation}
\gamma (q)+q_{i}(p_{1})\rightarrow q_{i}(p_{2})+g(p_{3}),\hspace{1in}(\gamma 
\overline{q}_{i}\rightarrow \overline{q}_{i}g),  \label{37}
\end{equation}
and photon-gluon fusion reactions 
\begin{equation}
\gamma (q)+g(p_{1})\rightarrow q_{i}(p_{2})+\overline{q}_{i}(p_{3}).
\label{38}
\end{equation}
In this article we consider the inclusive cross section difference in the
photon-proton collision, namely
\begin{equation}
\Delta _{M}=\frac{d\sigma}{dp_{T}^{2}dy}(\gamma p\rightarrow M^{+}X)-
\frac{d\sigma}{dp_{T}^{2}dy}(\gamma p\rightarrow M^{-}X)\equiv \Sigma_{M^{+}}-
\Sigma_{M^{-}}.
\end{equation} 
The LT subprocess which dominates in this difference is 
$\gamma q\rightarrow gq$ with $q\rightarrow M$. Its cross section at the tree
level is well known,
\begin{equation}
\frac{d{\hat\sigma}^{LT}}{d{\hat t}}=-\frac{8\pi \alpha_{E}e_{q}^{2}}{3{\hat s}^{2}}
\left [\alpha_{S}({\hat s})\frac{{\hat t}}{{\hat s}}+\alpha_{S}(-{\hat t})\frac{{\hat s}}{{\hat t}} \right],
\end{equation}
where the Mandelstam invariants of the subprocess are 
$${\hat s} =(q+p_{1})^{2},~~ {\hat t} =(q-p_{2})^{2},~~ {\hat u} =(q-p_{3})^{2}.$$
In (42), the running coupling constant $\alpha_{S}$ is evaluated at momentum
scales ${\hat s}$ and $\mid {\hat t}\mid $, which are equal to off-shell momenta 
carried by the virtual quark propagators in the corresponding Feynman diagrams 
of the leading twist subprocess $\gamma q\rightarrow gq$.

Many other subprocesses contribute to the meson photoproduction, among them
i) $\gamma q\rightarrow gq$ with $g\rightarrow M$, ii) $\gamma {\bar q}\rightarrow g{\bar q}$
with ${\bar q}\rightarrow M$, iii)$\gamma {\bar q}\rightarrow g{\bar q}$ with
$g\rightarrow M$, iv)$\gamma g\rightarrow q{\bar q}$ with $q\rightarrow M$ or
${\bar q}\rightarrow M$. Considering the cross section difference 
$\Delta_{M}$ we not only reduce the number of subprocesses contributing to
$\Delta_{M}$, because the subprocesses involving gluons or antiquarks contribute 
equally to $\Sigma_{M^{+}}^{LT}$ and $\Sigma_{M^{-}}^{LT}$ and cancel in
$\Delta_{M}^{LT}$, but also solve two other important problems. The first one 
is the next-to-leading order correction to the meson photoproduction cross 
section calculated in [18]. In this paper the authors have investigated 
the ratio
$$C^{\pi^{+}-\pi^{-}}=\frac{d\sigma^{HO}(\gamma p\rightarrow\pi^{+}X)/
d{\bf p}_{T}dy-d\sigma^{HO}(\gamma p\rightarrow \pi^{-}X)/d{\bf p}_{T}dy}
{d\sigma^{Born}(\gamma p\rightarrow \pi^{+}X)/d{\bf p}_{T}dy-d\sigma^{Born}
(\gamma p\rightarrow \pi^{-}X)/d{\bf p}_{T}dy}=K-1,$$
for the cross section difference as a function of $p_{T}$ at $\sqrt{s}=14.1~GeV$
and $y=0.5$. They have found that this ratio is negative and almost constant with 
$p_{T}$ $(p_{T}=2-6~GeV/c)$. This means that the K-factor for the cross section
difference is less than 1. In other words, using the LT cross section (42) for 
calculations of $R_{M}=\mid \Delta_{M}^{HT}/\Delta_{M}^{LT} \mid$ at the same 
or slightly different kinematic regimes, we only underestimate 
the ratio $R_{M}$ and related quantities and give lower bounds for them. 
The second problem solved by our choice of $\Delta_{M}$ is a contribution to 
the photoproduction cross section originating from the photon's quark and 
gluon content. It is well known that the photoproduction process 
$\gamma p\rightarrow h+X$ may proceed via two distinct mechanisms; 
photon can interact either directly with the hadron's partons (direct 
photoproduction), or via its quark and gluon content (resolved photoproduction). 
As was demonstrated in [18], the contribution from the resolved photoproduction 
almost completely cancel in the $\pi^{+}-\pi^{-}$ difference. These results
obtained in [18] for pions at certain kinematic domain seemingly are valid also
for other light mesons at the same or slightly different kinematic conditions.

Then the leading twist contribution to the single meson photoproduction in
$\gamma p\rightarrow MX$ is given by the expression,
\begin{equation}
\frac{d\sigma^{LT}}{dp_{T}^{2}dy}=\sum_{q}\int_{x_{min}}^{1}\frac{dxq_{p}
(x,-{\hat t})D_{M/q}(z,-{\hat t})}{z}\frac{d{\hat \sigma}^{LT}}{d{\hat t}},
\end{equation}
where
\begin{equation}
z=\frac{p_{T}e^{-y}}{x\sqrt{s}}+\frac{p_{T}e^{y}}{\sqrt{s}}, 
~~~x_{min}=\frac{p_{T}e^{-y}}{\sqrt{s}-p_{T}e^{y}}.
\end{equation}
In (43), $q_{p}(x,-{\hat t})$ and $D_{M/q}(z,-{\hat t})$ are a quark $q$ distribution
and fragmentation functions, respectively. The subprocess invariants ${\hat s}$,
${\hat t}$, ${\hat u}$ in (43) are functions of $s,~p_{T},~y,$
\begin{equation}
{\hat s}=xs, ~~{\hat t}=-\frac{p_{T}\sqrt{s}e^{-y}}{z},~~{\hat u}=-\frac{xp_{T}
\sqrt{s}e^{y}}{z}.
\end{equation}
Eq.(43) together with (37) for the HT contributions will be applied
in the next section for numerical calculations.

\section{NUMERICAL RESULTS}

In this section we compute the $\gamma p\rightarrow M^{+}X$ and $\gamma p
\rightarrow M^{-}X$ inclusive cross sections $\Sigma_{M^{+}},~\Sigma_{M^{-}}$,
as well as the difference $\Delta_{M}=\Sigma_{M^{+}}-\Sigma_{M^{-}}$ by taking
into account the dominant LT ($\gamma q\rightarrow gq$ with $q\rightarrow M$), 
and HT ($\gamma q \rightarrow Mq$) contributions to the inclusive photoproduction.
Only the HT cross section of $K^{-}$ photoproduction is calculated using the 
proton $s$ and ${\bar u}$ quarks induced subprocesses, which contribute at 
the same order. Our calculations are performed for $M=\pi~ ,K,~\rho$ at 
$\sqrt{s}=14.1~GeV,~25~GeV$.

In this work, for quark distribution functions, we borrow the leading order 
parametrization of Owens [19]. This parametrization is suitable for our purposes,
because the HT mechanism probe the quark distribution functions at\\
$z^{\ast}=p_{T}\exp{(-y)}/\sqrt{s} - p_{T}\exp{(y)}$, which for chosen process's 
parameters is always more than $0.01$. The same is also true for $x_{min}$ in
the LT cross section (43). That is, kinematical conditions allow us
to avoid the region of small $x\leq 0.01$, where Owen's parametrization may give
incorrect results. The same reason will enable us to compute $\Sigma_{M}^{LT}$
ignoring a contribution from the leading twist subprocess $\gamma g\rightarrow
q{\bar q}$, which otherwise may be considerable.

The quark fragmentation functions are taken from [20]. Recently, in [21], a new 
set of fragmentation functions for charged pions and kaons, both at leading
and next-to-leading order, have been presented. These functions give 
$D_{q}^{M^{+}+M^{-}}(x,Q^{2})$, but not $D_{q}^{M^{\pm}}(x,Q^{2})$. Therefore, we
cannot apply them in our calculations. 

The other problem is a choice of the QCD scale parameter $\Lambda$ and number
of active quark flavors $n_{f}$. The HT subprocesses probe the meson w.f. over
a large range of $Q^{2}$, $Q^{2}$ being equal to ${\hat s}$ or $-{\hat u}$. It is 
easy to find that $-{\hat u}_{min}>4.04~GeV^{2}$, while ${\hat s}_{min}>16~GeV^{2}$.
For momentum scales ${\hat s}, -{\hat t}$ used in (42) as arguments of $\alpha_{S}$
in the LT cross section we get
$$-{\hat t}_{min}>6~GeV^{2},~~~ {\hat s}_{min}>16~GeV^{2}.$$

In other kinematic domains these scales take essentially larger values. 
Taking into account these facts we find it reasonable to assign 
$\Lambda=0.1~GeV, n_{f}=5$ throughout in this section.

Results of our numerical calculations are plotted in $Figs.2-8$. First of all,
it is interesting to compare the resummed HT cross sections with the ones obtained
in the framework of the frozen coupling approximation. In $Fig.2$, the 
ratio $r_{M}=(\Sigma_{M}^{HT})^{res}/(\Sigma_{M}^{HT})^{o}$ for negatively
charged particles $(\pi^{-},K^{-})$ is shown. In the computing of 
$(\Sigma_{M}^{HT})^{o}$ we have neglected the meson's w.f. dependence on the
scale ${\hat Q}^{2}$. Let us emphasize that for the kaon we have used the frozen 
coupling version of our expression (9), but not the Bagger-Gunion formula
from [6], which is incorrect in that case. 

As is seen from $Fig.2(a)$, $r_{\pi^{-}}\simeq 1$ almost for all 
$p_{T}$, whereas $r_{K^{-}}$ falls from $r_{K^{-}}\simeq 2.75$
at $p_{T}=2~GeV/c,~ y=0$ until $r_{K^{-}} \simeq 2.13$ at $p_{T}=11~GeV/c,~y=0$
and from $r_{K^{-}}\simeq 2.71$ at $p_{T}=2~GeV/c,~y=0.5$ till 
$r_{K^{-}} \simeq1.5$ at $p_{T}=11~GeV/c,~y=0.5$. For $K^{-}$ this ratio 
demonstrates also a sharp dependence on $y$ at fixed $\sqrt{s},~p_{T}$ $(Fig.2(b))$. 

In all of the following figures we have used the resummed expression for the 
HT cross section. In $Fig.3$ the ratio $R_{M}=\mid \Delta_{M}^{HT}/
\Delta_{M}^{LT} \mid$ is depicted. For all particles the LT cross section 
difference is positive $\Delta_{M}^{LT}>0$, since 
$\Sigma_{M^{+}}^{LT}\sim u_{p}(x,-{\hat t})e_{u}^{2}$, while 
$\Sigma_{M^{-}}^{LT}\sim d_{p}(x,-{\hat t})e_{d}^{2}$. The smaller quark charge
$e_{d}$ and the smaller distribution function $d_{p}$ both suppress 
$\Sigma_{M^{-}}^{LT}$ [6]. The HT cross section difference may change
sign at small $p_{T}$ and become negative $\Delta_{M}^{HT}<0$. For example,
$\Delta_{\pi^{-}}^{HT}<0$ at $2~GeV/c\leq p_{T}\leq 11~GeV/c$ for 
$\sqrt{s}=25~GeV,~ y=0$ and at $2~GeV/c\leq p_{T}\leq 9~GeV/c$ for 
$\sqrt{s}=25~GeV,~y=0.5$.
Only at the phase-space boundary $p_{T}>11~GeV/c$ in the first case or at
$p_{T}>9~GeV/c$ in the second one $\Sigma_{\pi^{+}}^{HT}>\Sigma_{\pi^{-}}^{HT}$.
Therefore, we plot the absolute value of $R_{M}$. The similar picture 
has been also found for other mesons. 

As is seen from $Figs.3(a),(b)$ for pion and kaon the HT contribution is comparable
with the LT one only at $p_{T}\leq 3~GeV/c$. We do not find a considerable
and stable growth of HT contributions at large values of $p_{T}$ for all
the cross section differences $\Delta_{M}^{HT}~(M=\pi ,K)$, as well as, for all
$\Sigma_{M}^{HT}$. Thus, $\Sigma_{K^{-}}^{HT}/\Sigma_{K^{-}}^{LT}$
is small at high $p_{T}$ for different $\sqrt{s}$  $(Fig.4(a))$. At the same 
time $\Sigma_{K^{+}}^{HT}/\Sigma_{K^{+}}^{LT}$ is a rising function of $p_{T}$
for $p_{T}>4~GeV/c~ (\sqrt{s}=14.1~GeV)$ and $p_{T}>7~GeV/c~ (\sqrt{s}=25~GeV)$.

The cross section differences $\Delta_{M}^{LT}$ and $\Delta_{M}^{tot}=
\Delta_{M}^{LT}+\Delta_{M}^{HT}$ as functions of $p_{T}$ are shown in $Fig.5$.
For the pion the total cross section difference $\Delta_{\pi}^{tot}$ in the 
region of small $p_{T}$ is smaller than $\Delta_{\pi}^{LT}$ due to 
$\Delta_{\pi}^{HT}<0$ in this region $(Fig.5(a))$. But for kaon 
$\Delta_{K}^{tot}>\Delta_{K}^{LT}$ in the same kinematic domain. For both 
mesons the difference between $\Delta_{M}^{tot}$ and $\Delta_{M}^{LT}$ cross
sections is small.

The rapidity dependence of $R_{M}$ at $\sqrt{s}=25~GeV, p_{T}=3~GeV/c$ plotted in
$Fig.3(c)$ illustrates not only the tendency of the HT contributions to be
enhanced in the region of negative rapidity, but also reveals an interesting
feature of the HT terms; as is seen from $Fig.3(c)$ the ratio $R_{M}$ is an
oscillating function of the rapidity. This property of the HT terms may have
important phenomenological consequences. In fact, in $Fig.6$ we have depicted
$\Delta_{M}^{tot}$ and $\Delta_{M}^{LT}$ versus rapidity. In both cases, owing
to observed property of $\Delta_{M}^{HT}(y)$, in certain domains of the 
rapidity interval $-2\leq y\leq 2.105$ the total cross section difference is more
than $\Delta_{M}^{LT}$ and in some ones less than $\Delta_{M}^{LT}$. In the case
of the kaon photoproduction 
\begin{eqnarray*}
&\Delta_{K}^{tot}>\Delta_{K}^{LT},~~ for~~ -2\leq y\leq 0.3~~ and~~ 1.8\leq y\leq 2.105,\\
&\Delta_{K}^{tot}<\Delta_{K}^{LT},~~ for~~ 0.3\leq y\leq 1.8.
\end{eqnarray*}

The properties of the HT terms found in the pion and kaon photoproduction
processes persist also in the $\rho$-meson photoproduction. But now the HT
contributions change the whole picture of the process arising from the ordinary 
LT calculations. Thus, as in the case of the pion photoproduction, 
the HT terms are enhanced relative to the leading ones and 
$\Delta_{\rho}^{HT}<0$ almost for all $p_{T}$. But now 
$\mid \Delta_{\rho}^{HT}\mid$ takes such large values that it even changes the
sign of the total cross section difference. That is, if in accordance with the
LT estimations $\Sigma_{\rho^{+}}^{tot}>\Sigma_{\rho^{-}}^{tot}$ 
must be valid for all $p_{T}$, for $p_{T}<p_{T}^{c}$ we find $\Sigma_{\rho^{+}}^{tot}
<\Sigma_{\rho^{-}}^{tot}$. The value of $p_{T}^{c}$ depends on the process 
parameters, as well as on the $\rho$-meson w.f. used in calculations. At $p_{t}
\approx p_{T}^{c}$ we have $\Sigma_{\rho^{+}}^{tot}\approx \Sigma_{\rho^{-}}^
{tot}$.

Our results are shown in $Fig.7$. For the parameters indicated in the 
figure a critical value of $p_{T}$ is: $p_{T1}^{c}\simeq 5.05~GeV/c$ 
for CZ w.f., and $p_{T2}^{c}\simeq 6.25~GeV/c$ for BB w.f. In all 
kinematic domains the HT contributions found using BB w.f. exceed the ones 
obtained by applying CZ w.f., that is, $\mid \Delta_{\rho}^{HT}(BB)\mid>\mid \Delta_{\rho}^{HT}(CZ)\mid $. 
For example, the ratio $\mid \Delta_{\rho}^{HT}(BB)/\Delta_{\rho}^{HT}(CZ)\mid $ equals
to $2.39$ at $\sqrt{s}=25~GeV,~ p_{T}=5~GeV/c,~ y=0$, or to $2.63$ at $\sqrt{s}=25~GeV,~
p_{T}=3~GeV/c,~ y=-1$. Our results confirm the conclusion made by the authors in [13] 
concerning a possibility of increasing the rate of the production of 
transversely polarized $\rho$-meson. Similar pictures persist in $Fig.8$, where 
$\Delta_{\rho}^{LT}$ and $\Delta_{\rho}^{tot}$ are depicted as functions of the
rapidity y. In $Fig.8(a)$, for the process parameters $\sqrt{s}=25~GeV, 
p_{T}=3~GeV/C$ we have: in domain $I'~(-1.74\leq y\leq 1.3)$ the total 
cross section difference for CZ w.f. is negative, in 
$I~(-1.5\leq y\leq 1.5)$~-~ $\Delta_{\rho}^{tot}(BB)<0$. In $Fig.8(b)$ the same 
is shown for $\sqrt{s}=25~GeV,~ p_{T}=5~GeV/c$.
In two other regions lying outside of $I(I')$ the $\Sigma_{\rho^{+}}^{tot}$
exceeds $\Sigma_{\rho^{-}}^{tot}$ (for $-2\leq y\leq -1.5$, $\Sigma_{\rho}^{tot}
(BB)$ and for $-2\leq y\leq -1.74$, $\Sigma_{\rho}^{tot}(CZ)$ are negligible
and are not shown).

It is worth noticing that in [6] the authors considered the $\rho$-meson 
photoproduction at the same process's parameters and predicted $\Sigma_{\rho}^{tot}<0$
at $p_{T}\leq 3~GeV/c$, but could not find similar effects for
$\Sigma_{\rho}^{tot}$ in dependence on the rapidity. Our investigations prove
that $\Sigma_{\rho}^{tot}<0$ at $p_{T}<p_{T}^{c}$ and $p_{T}^{c}$ well into deep
perturbative domain. We have also demonstrated that the same phenomenon exists
for $y_{1}^{c}\leq y\leq y_{2}^{c}$.

\section {CONCLUDING REMARKS}

In this work we have calculated the single meson inclusive photoproduction via
higher twist mechanism and obtained the expressions for the subprocess 
$\gamma q\rightarrow Mq$ cross section for mesons with both symmetric and
non-symmetric wave functions. For the calculation of the cross section we have 
applied the running coupling constant method and revealed IR renormalon poles
in the cross section expression. IR renormalon induced divergences have been
regularized by means of the principal value prescription and the resummed
expression (the Borel sum) for the higher twist cross section has been found.
Phenomenological effects of the obtained results have been discussed.

Summing up we can state that:\\
i) for mesons with non-symmetric w.f. in the framework of the frozen coupling 
approximation the higher twist subprocess cross section cannot be normalized
in terms of a meson electromagnetic form factor;\\
ii) in the context of the running coupling constant method the HT subprocess
cross section cannot be normalized in terms of meson's elm form factor
neither for mesons with symmetric w.f. nor for non-symmetric ones;\\
iii) the resummed HT cross section differs from that found using the frozen
coupling approximation, in some cases, considerably;\\
iv) HT contributions to the single meson photoproduction cross section have
important phenomenological consequences, specially in the case of $\rho$-meson
photoproduction. In this process the HT contributions wash the LT results off,
changing the LT predictions.
\vspace{5mm}

{\Large \bf ACKNOWLEDGMENTS}\vspace{5mm}\\
The author would like to thank the International Centre for Theoretical
Physics, Trieste, for hospitality and Prof. S. Randjbar-Daemi for his 
interest in this work.\newpage

{\Large \bf REFERENCES}\vspace{10mm}\\
{\bf 1.} S.J.Brodsky, R.Blankenbecler and J.F.Gunion: Phys.Rev.D6, 2651 (1972);\\
~~~S.J.Brodsky and G.R.Farrar: Phys.Rev.Lett.31, 1153 (1973).\vspace{2mm}\\
{\bf 2.} G.P.Lepage and S.J.Brodsky: Phys.Rev.D22, 2157 (1980).\vspace{2mm}\\
{\bf 3.} V.L.Chernyak and A.R.Zhitnitsky: Phys.Rep.112, 173 (1980).\vspace{2mm}\\
{\bf 4.} S.S.Agaev: Phys.Lett. B283, 125(1992); Z.Phys.C-Particles and Fields57,
403 (1993);\\
E.L.Berger and S.J.Brodsky: Phys.Rev.Lett. 42, 940 (1979);\\
E.L.Berger: Z.Phys.C-Particles and Fields 4, 289 (1980).\vspace{2mm}\\
{\bf 5.} V.N.Baier and A.G.Grozin: Phys.Lett.B96, 181 (1980);\\
S.Gupta: Phys.Rev.D24, 1169 (1981).\vspace{2mm}\\
{\bf 6.} J.A.Bagger and J.F.Gunion: Phys.Rev.D25, 2287 (1982).\vspace{2mm}\\
{\bf 7.} J.A.Hassan and J.K.Storrow: Z.Phys.C-Particles and Fields14, 65 (1982).\vspace{2mm}\\
{\bf 8.} S.J.Brodsky, G.P.Lepage and P.B.Mackenzie: Phys.Rev.D28, 228 (1983).\vspace{2mm}\\
{\bf 9.} R.D.Field, R.Gupta, S.Otto and L.Chang: Nucl.Phys.B186, 429 (1981).\vspace{2mm}\\
{\bf 10.}S.S.Agaev: Phys.Lett.B360, 117 (1995); E. Phys.Lett.B369, 379 (1996);\\
S.S.Agaev: Mod.Phys.Lett.A10, 2009 (1995).\vspace{2mm}\\
{\bf 11.}S.S.Agaev: Mod.Phys.Lett.A11, 957 (1996); ICTP preprint IC/95/291, 
September 1995, hep-ph/9611215.\vspace{2mm}\\
{\bf 12.}A.H.Mueller: Nucl.Phys.B250, 327 (1985);\\
H.Contopanagos and G.Sterman: Nucl.Phys. B419, 77 (1994).\vspace{2mm}\\
{\bf 13.}P.Ball and V.M.Braun: Phys.Rev.D54, 2182 (1996).\vspace{2mm}\\
{\bf 14.}S.S.Agaev: Int.J.Mod.Phys.A8, 2605 (1993); Int.J.Mod.Phys.A9, 5077 (1994).
\vspace{2mm}\\
{\bf 15.}G.'t Hooft: In: The Whys of Subnuclear Physics, Proc.Int.School, Erice,
1977, ed. A.Zichichi, Plenum, New York, 1978;\\
V.I.Zakharov: Nucl.Phys.B385, 452 (1982).\vspace{2mm}\\
{\bf 16.}A.Erdelyi: Higher transcendental functions, v.2,
McGrow-Hill Book Company, New York, 1953.\vspace{2mm}\\
{\bf 17.}M.Neubert: Phys.Rev.D51, 5924 (1995);\\
P.Ball, M.Beneke and V.M.Braun: Nucl.Phys.B452, 563 (1995);\\
P.Ball, M.Beneke and V.M.Braun: Phys.Rev.D52, 3929 (1995);\\
M.Beneke and V.M.Braun: Phys.Lett.B348, 513 (1995);\\
C.N.Lovett-Turner and C.J.Maxwell: Nucl.Phys.B452, 188 (1995).\vspace{2mm}\\
{\bf 18.}P.Aurenche, R.Baier, A.Douiri, M.Fontannaz and D.Schiff: 
Nucl.Phys.B286, 553 (1987).\vspace{2mm}\\
{\bf 19.}J.F.Owens: Phys.Lett.B266, 126(1991).\vspace{2mm}\\
{\bf 20.}J.F.Owens: Phys.Rev.D19, 3279 (1979);\\
J.F.Owens, E.Reya and M.Gl\"{u}ck: Phys.Rev.D18, 1501 (1978).\vspace{2mm}\\
{\bf 21.}J.Binnewies, G.Kramer and B.A.Kniehl: DESY preprint DESY-95-048, 
March 1995.
\newpage

{\Large \bf FIGURE CAPTIONS}\vspace{10mm}\\

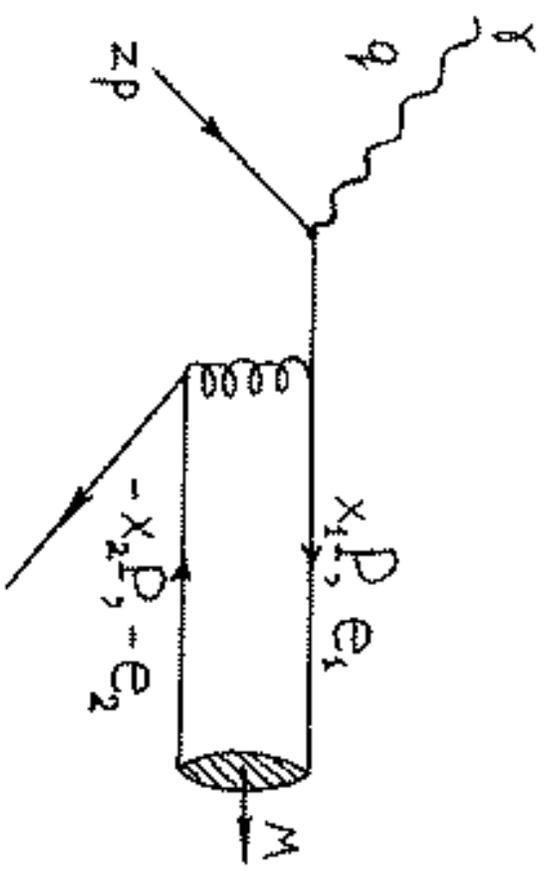
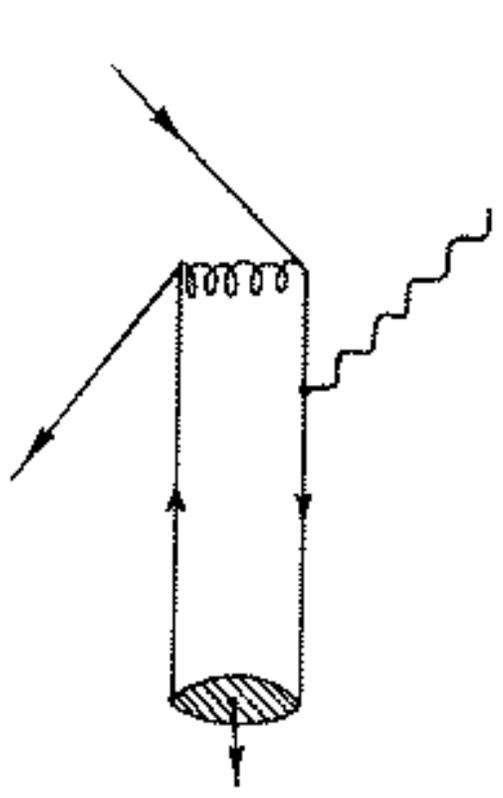
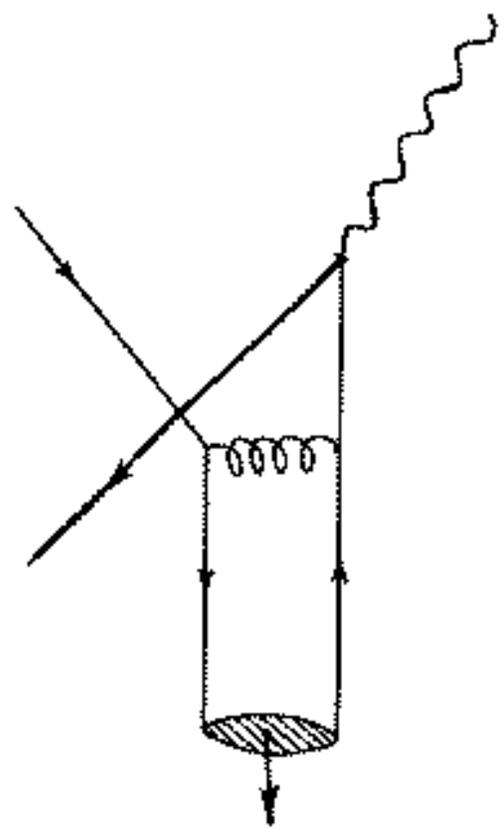
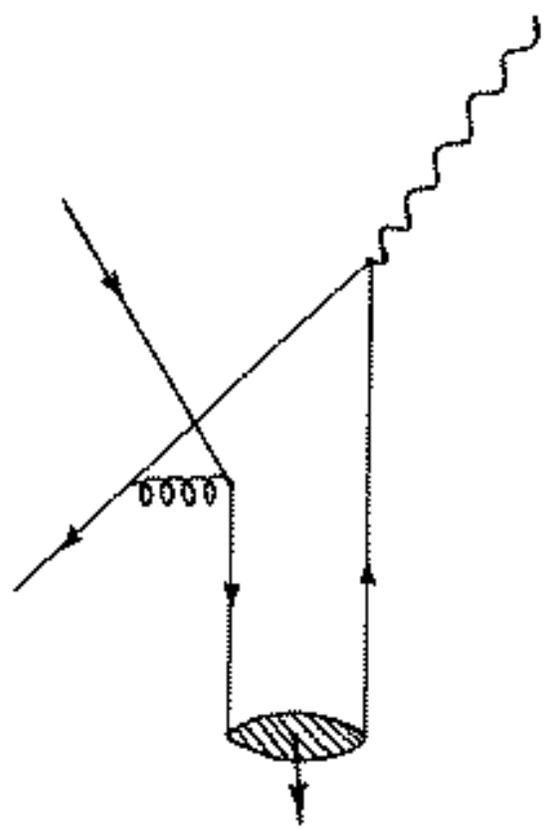
{\bf Fig.1} Feynman diagrams contributing to the higher twist subprocess $\gamma q
\rightarrow Mq$. Here ${\bf p}$ and ${\bf P}$ are the hadron h and meson M four 
momenta, respectively.
\vspace{5mm}

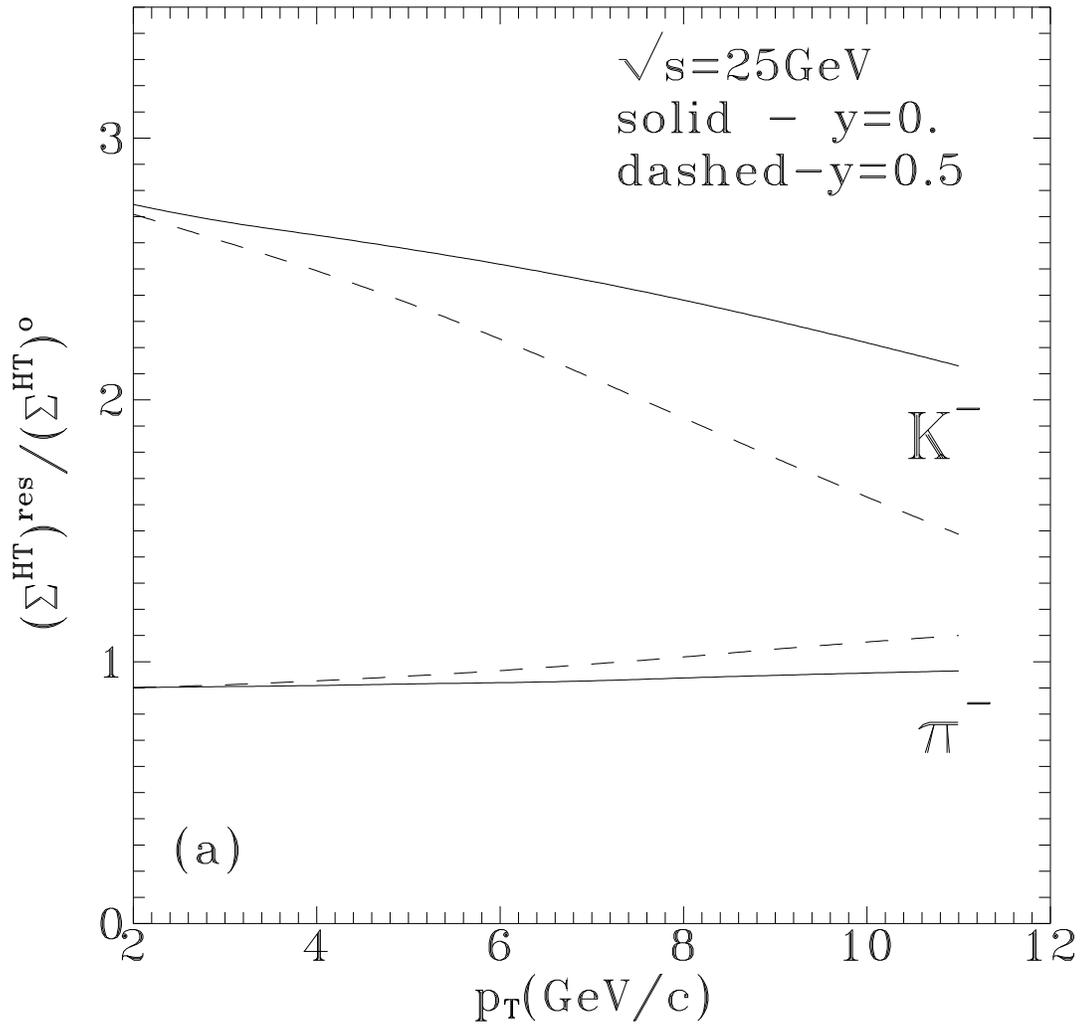
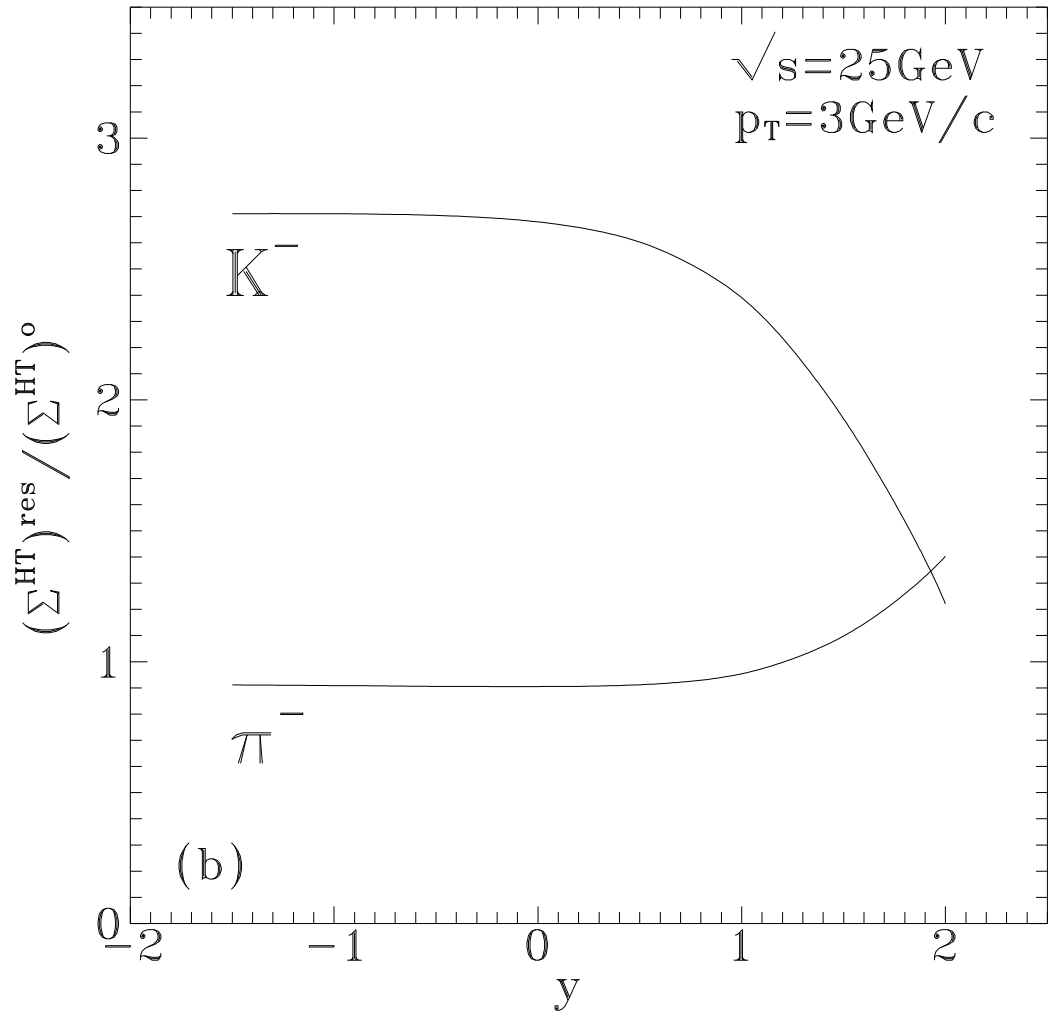
{\bf Fig.2} Ratio $r_{M}=(\Sigma_{M}^{HT})^{res}/(\Sigma_{M}^{HT})^{o}$, where
$(\Sigma_{M}^{HT})^{res}$ and $(\Sigma_{M}^{HT})^{o}$ are HT contributions
to the photoproduction cross section calculated using the running and frozen
coupling approximations, respectively. The ratio is depicted as a function
of $p_{T}$ a), and of the rapidity b).
\vspace{5mm}

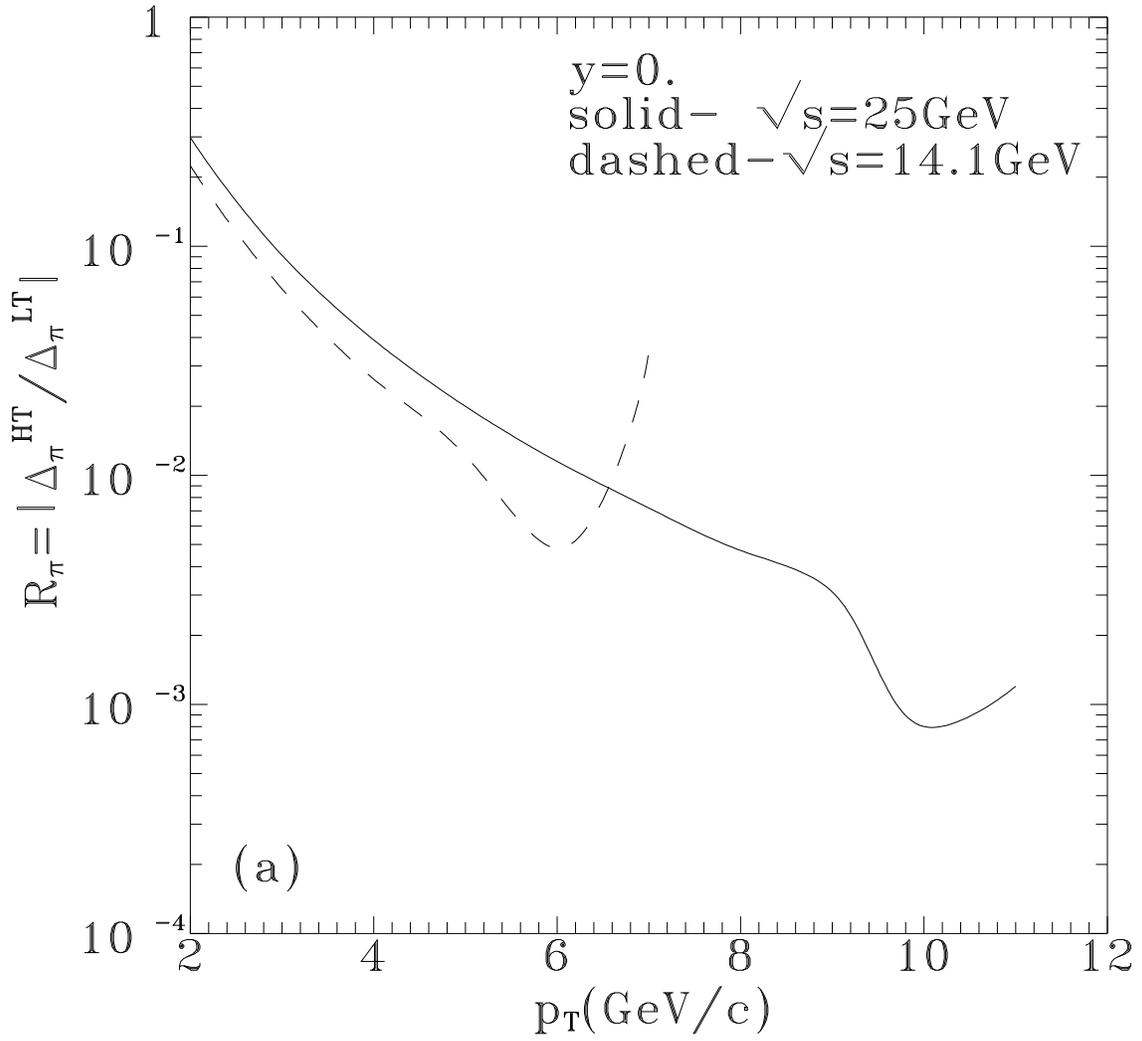
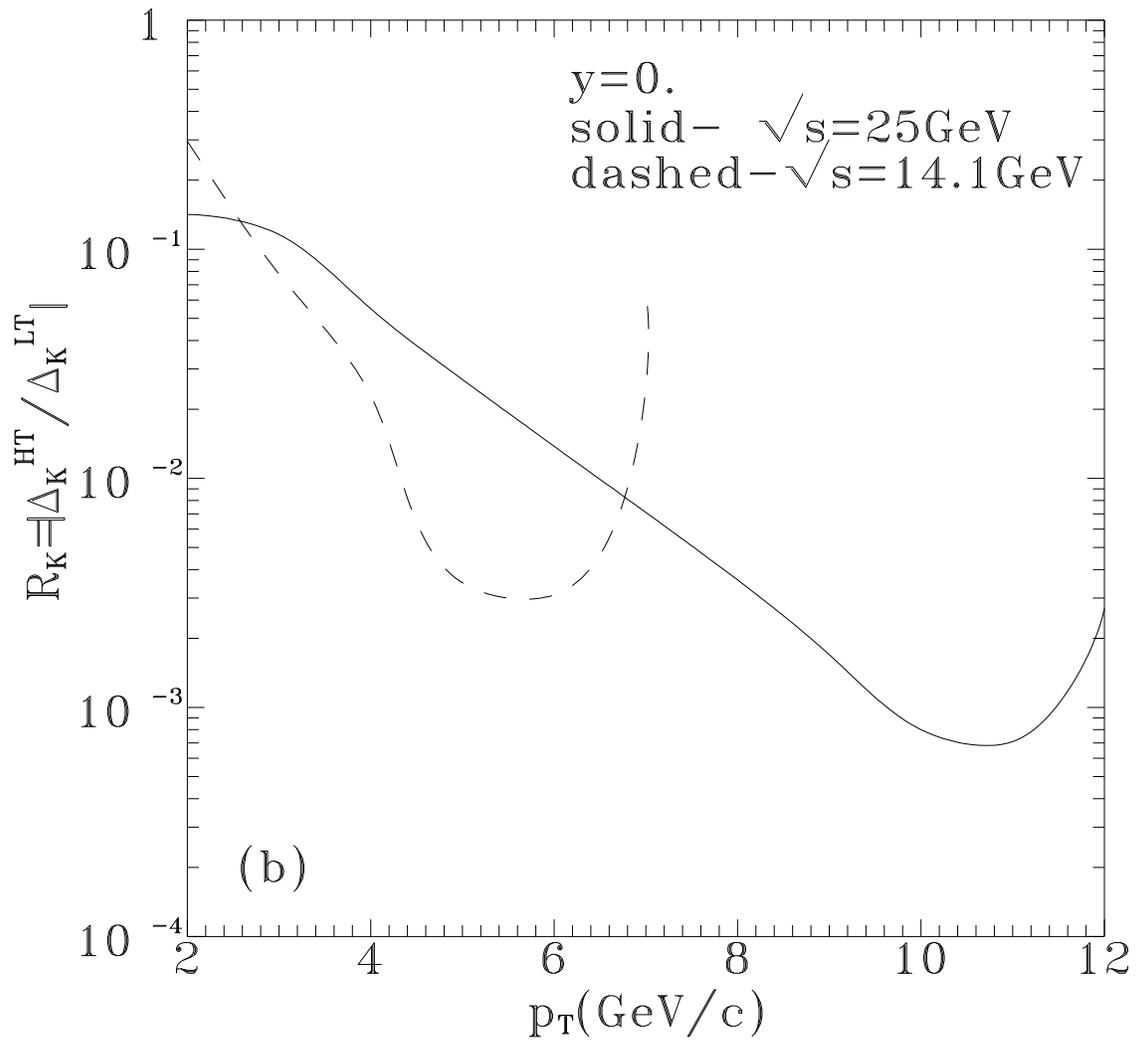
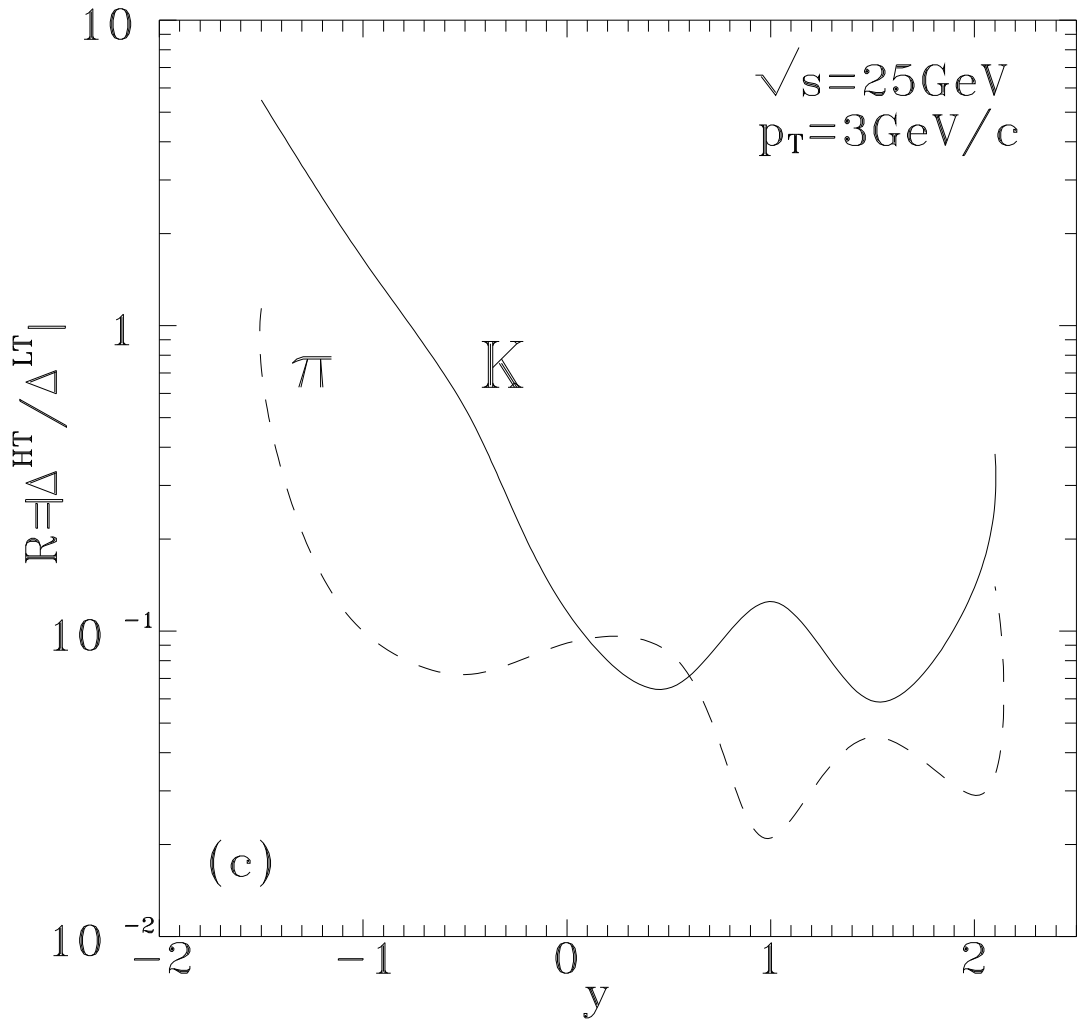
{\bf Fig.3} Ratio $R_{M}=\mid \Delta_{M}^{HT}/\Delta_{M}^{LT}\mid $ for the pion a),
and for the kaon b) at fixed rapidity y=0. In c) $R_{M}$ is plotted as a 
function of y for the pion (dashed curve) and for the kaon (solid curve).
\vspace{5mm}

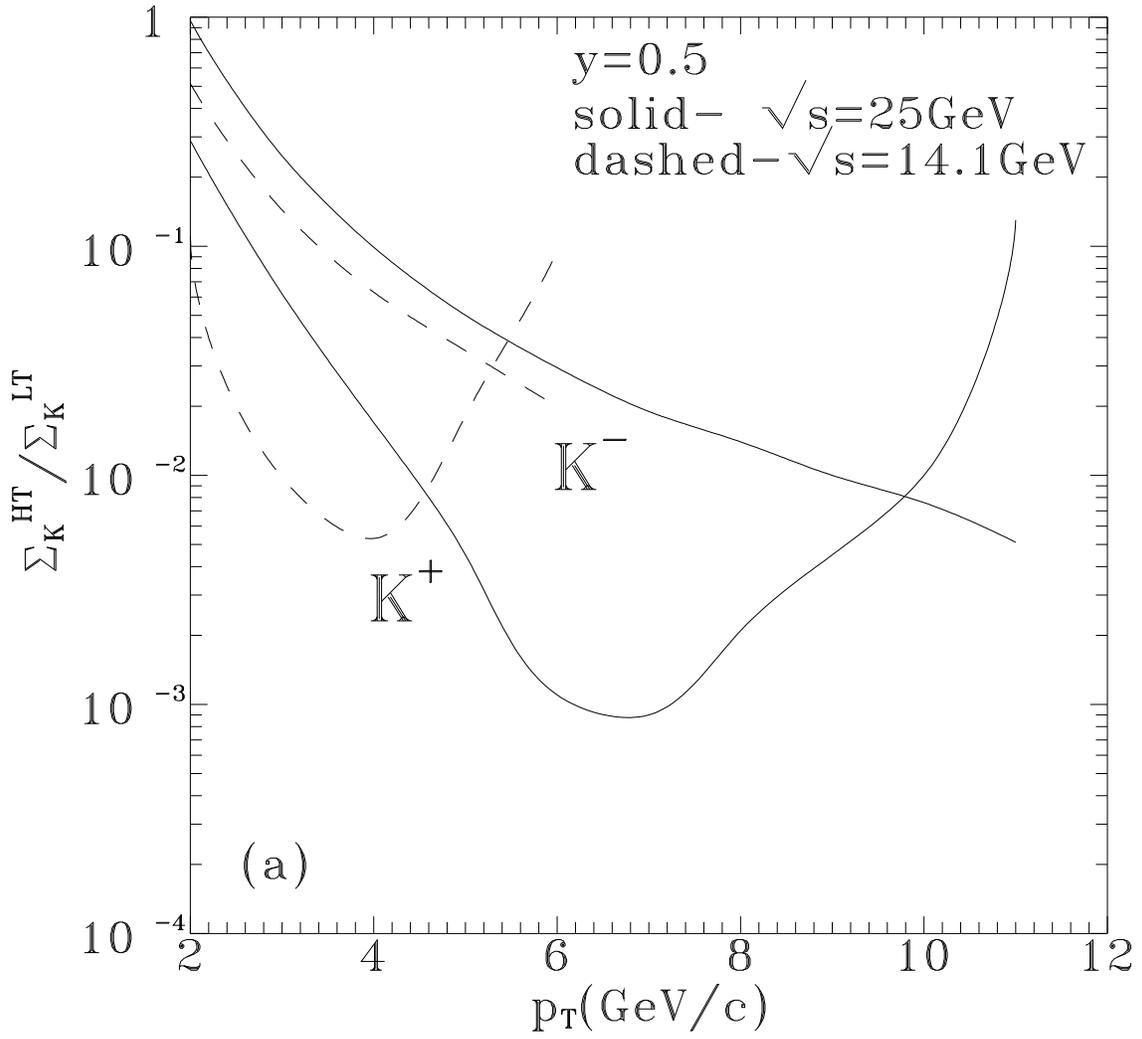
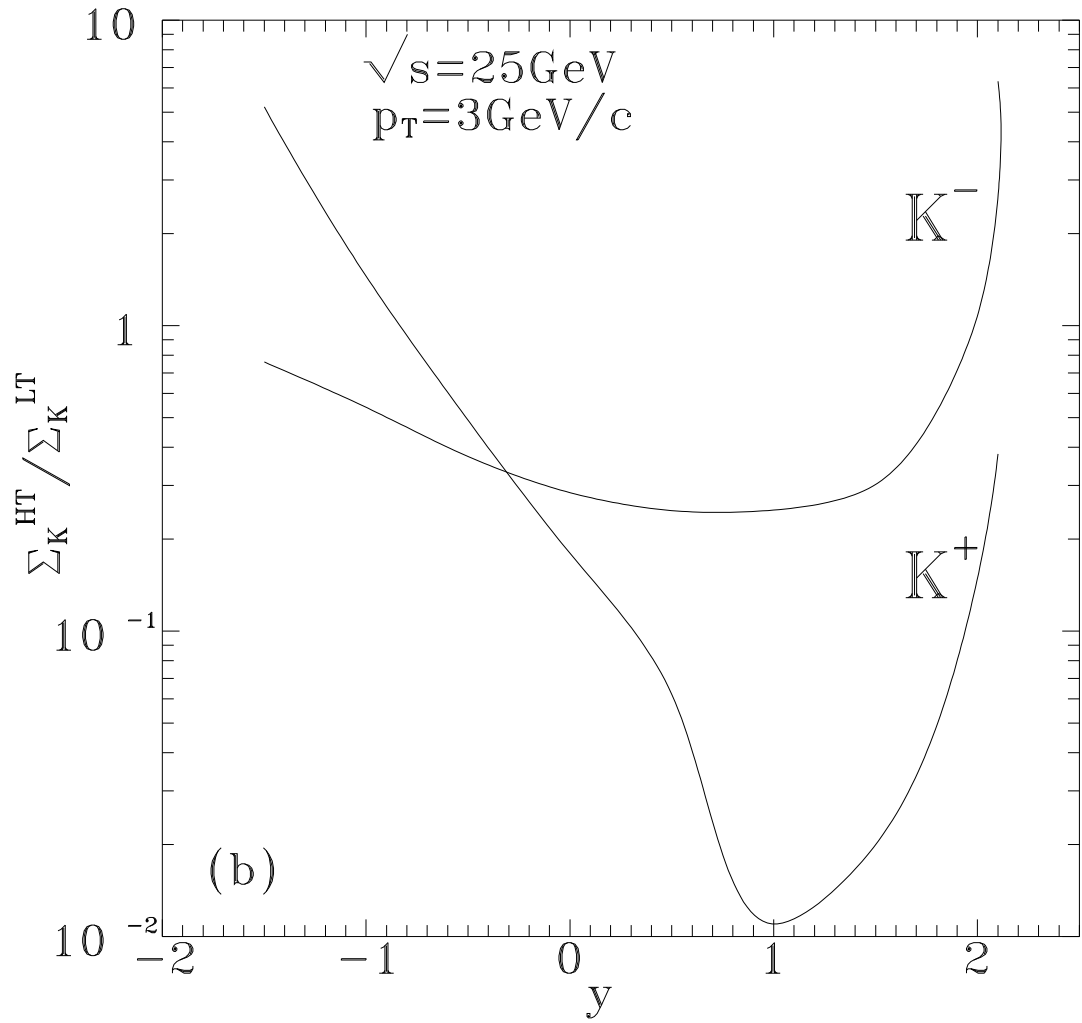
{\bf Fig.4} The dependence of the ratio $\Sigma_{K}^{HT}/\Sigma_{K}^{LT}$ for $K^{+}$
and $K^{-}$ on $p_{T}$ a) and on y b).
\vspace{5mm}

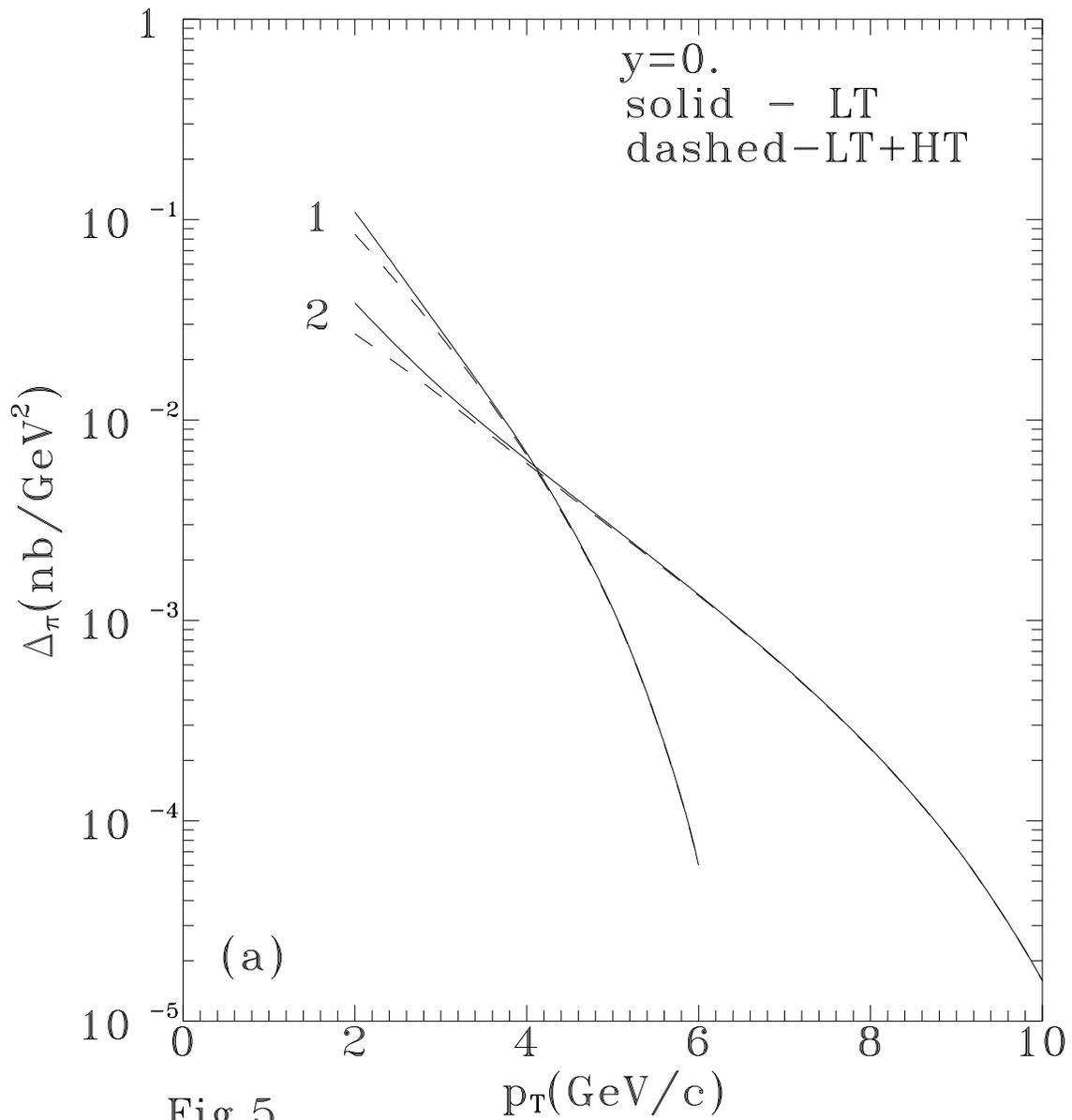
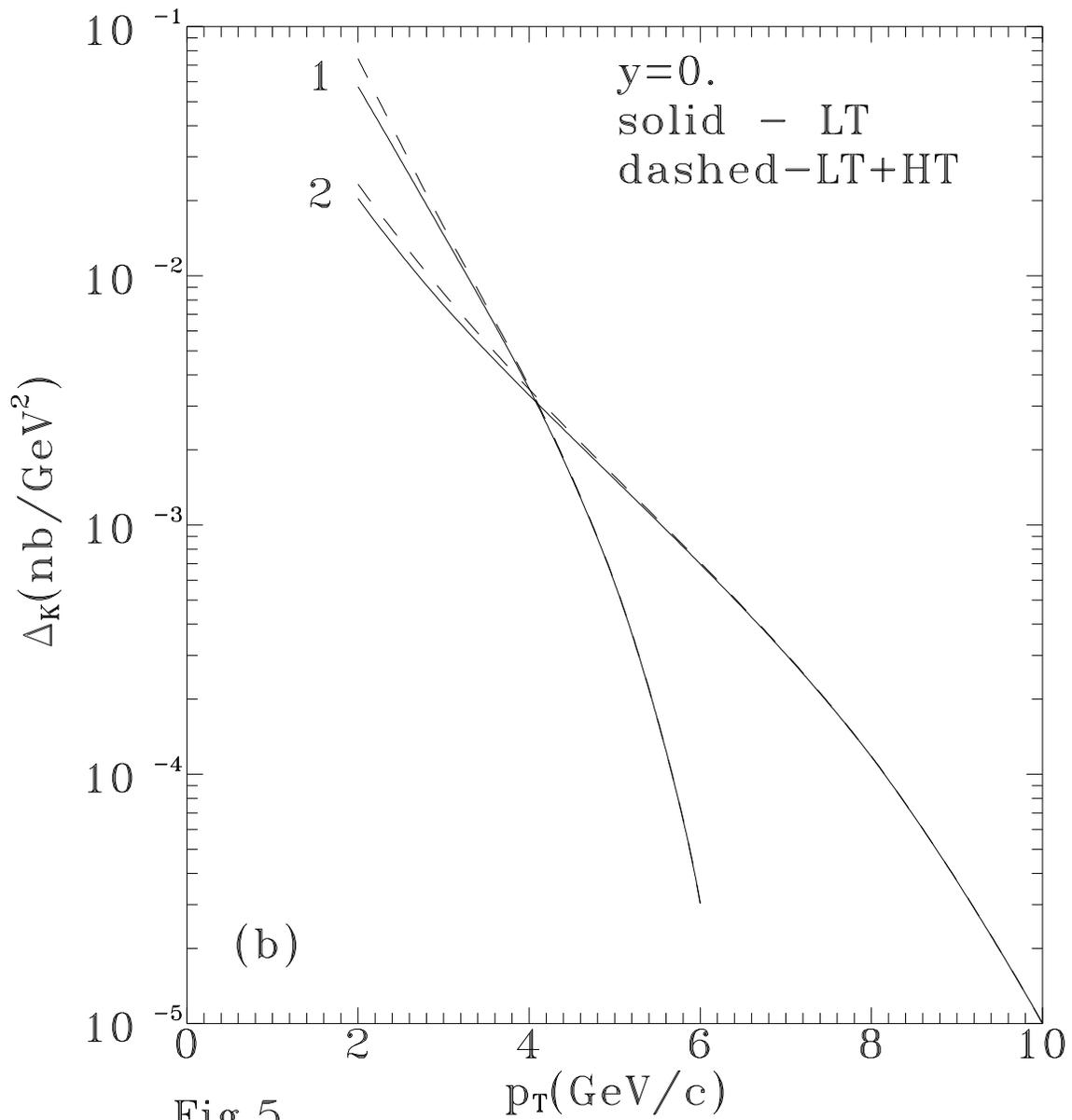
{\bf Fig.5} The cross section difference $\Delta_{M}$ is shown at fixed rapidity
for pions a), and for kaons b). For the curves {\bf 1} the process c.m. 
energy is $\sqrt{s}=14.1~GeV$, for the curves {\bf 2}~~-~~$\sqrt{s}=25~GeV$.
\vspace{5mm}

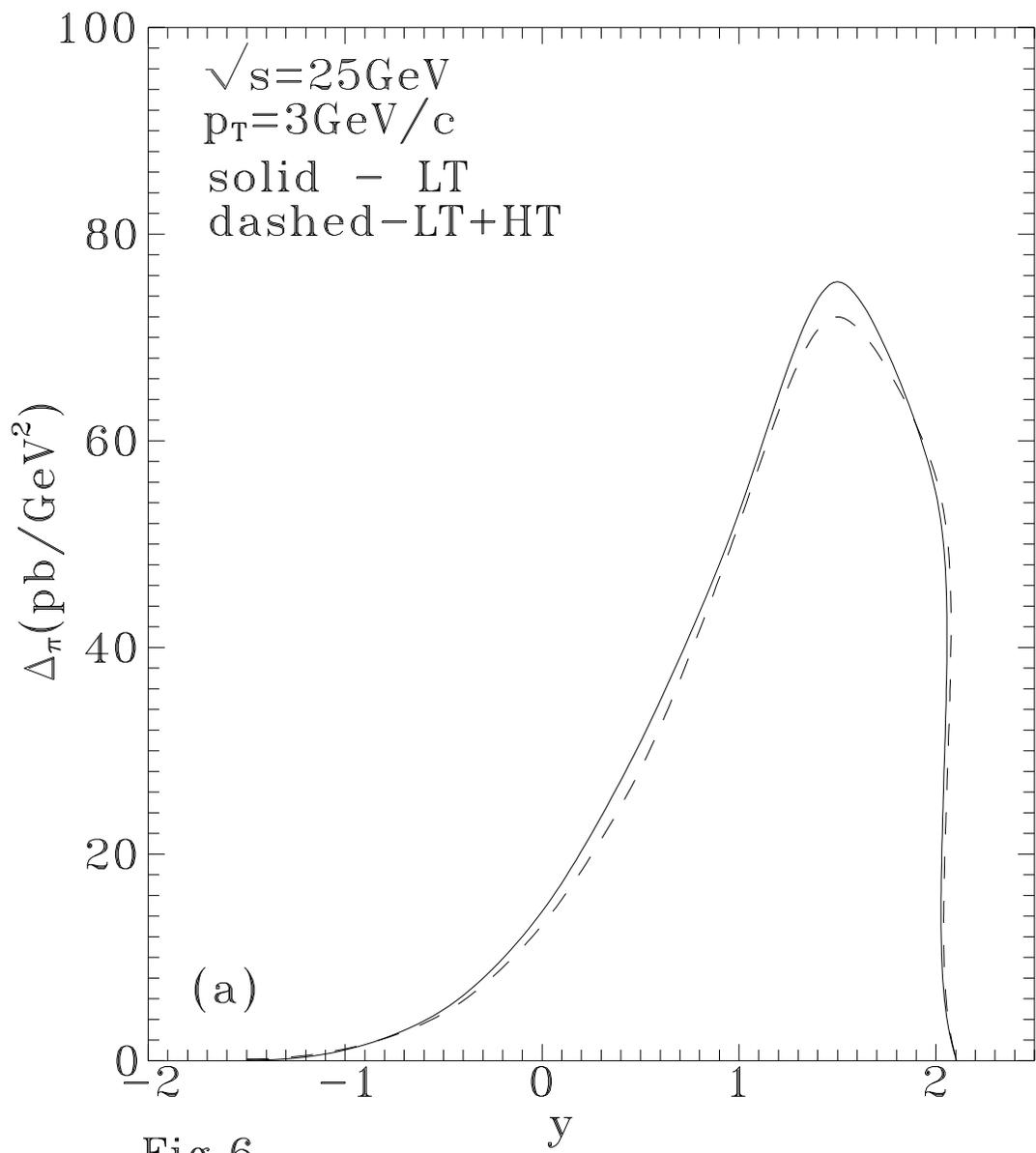
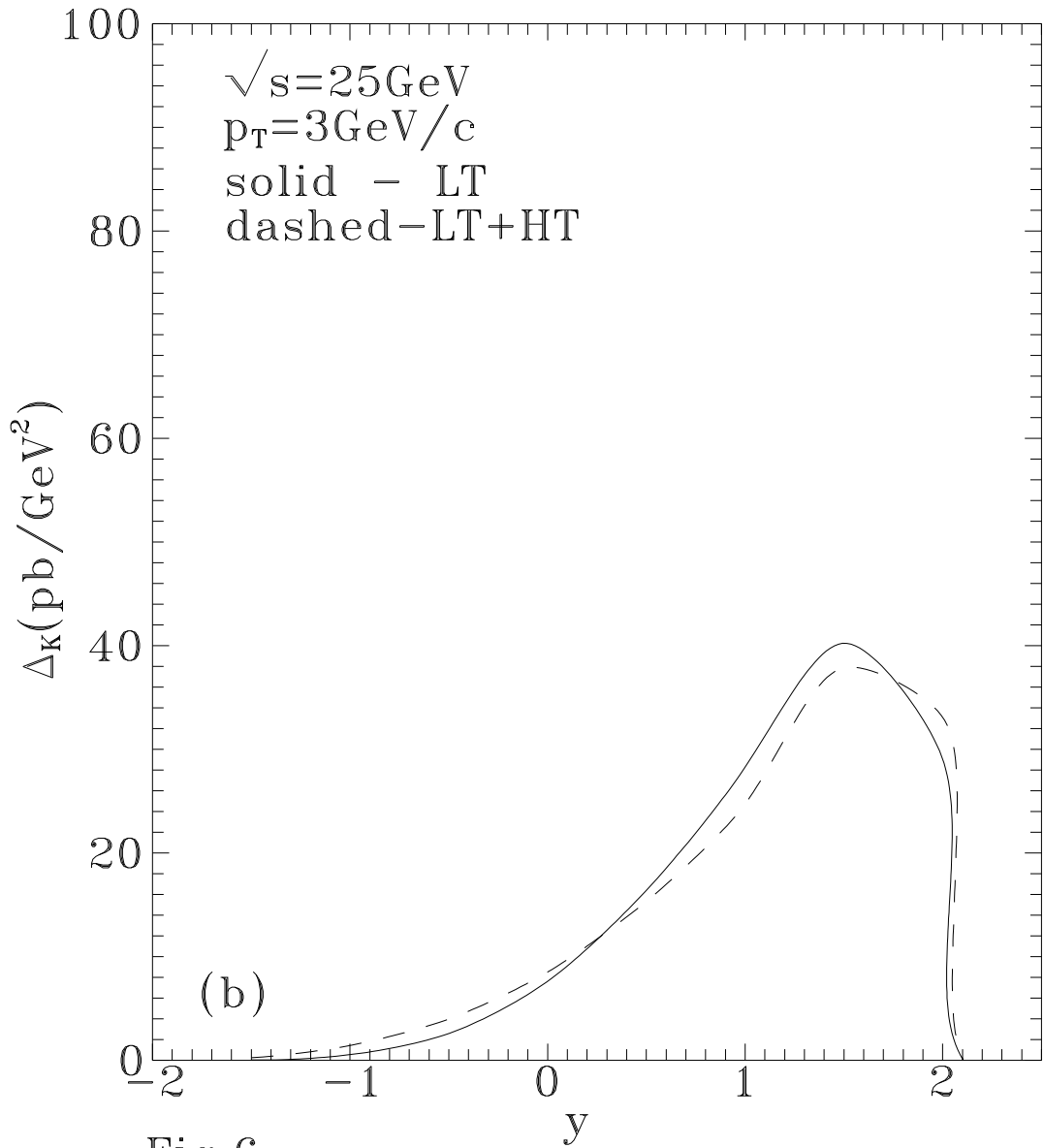
{\bf Fig.6} The cross section difference $\Delta_{M}$ as a function of the rapidity
for pions a); for kaons b).
\vspace{5mm}

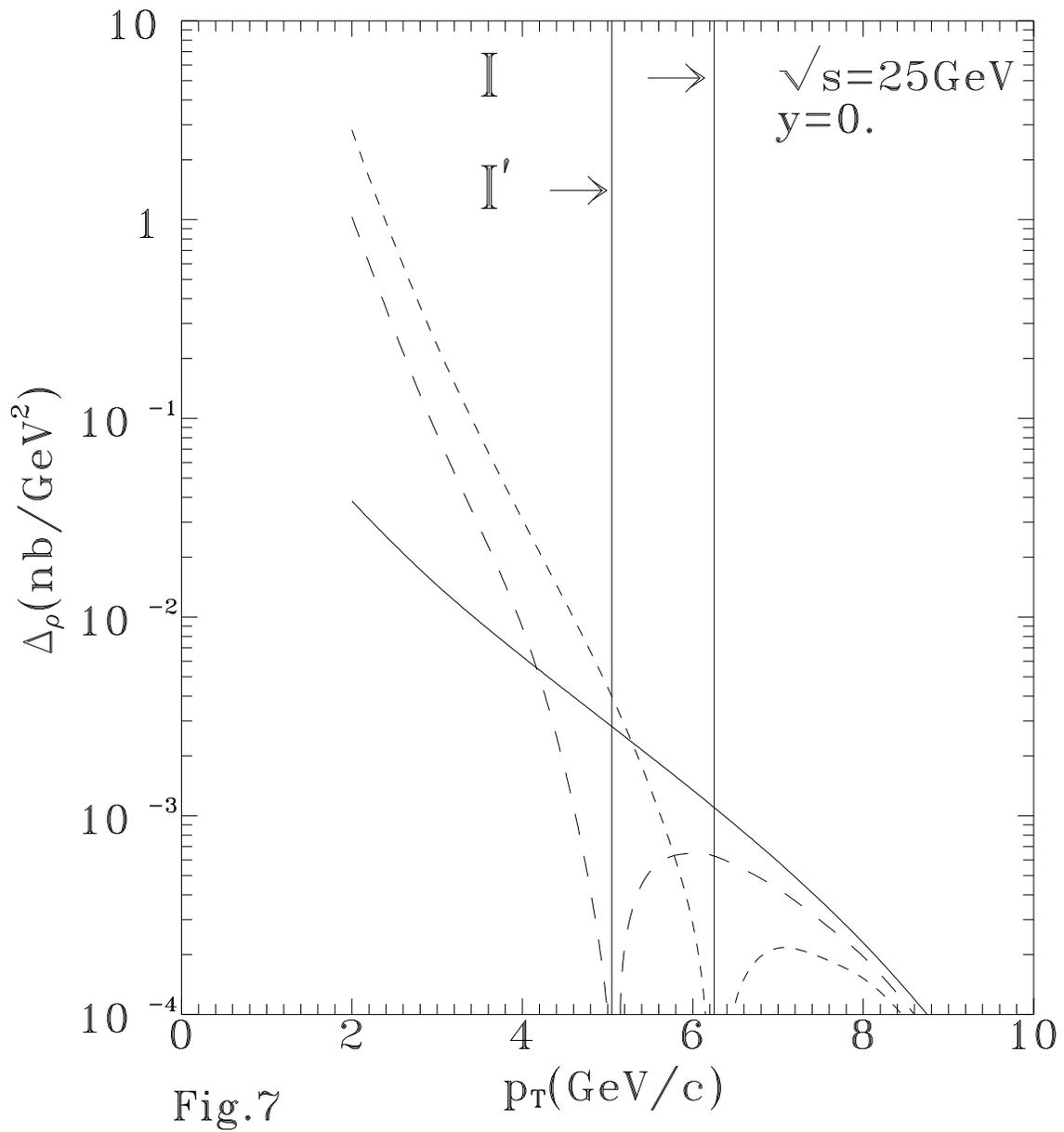
{\bf Fig.7} $\Delta_{\rho}$ for $\rho$-meson. The solid curve describes 
$\Delta_{\rho}^{LT}$, whereas the dashed curves correspond to 
$\Delta_{\rho}^{tot}$. The long-dashed curve has been obtained using the CZ
w.f., the short-dashed one- BB w.f. In the domains $I (BB~ w.f.)$ and 
$I' (CZ~ w.f.)$ the absolute value of $\mid \Sigma_{\rho}^{tot}\mid $ or
$\Sigma_{\rho^{-}}^{tot}-\Sigma_{\rho^{+}}^{tot}$ is plotted.
\vspace{5mm}

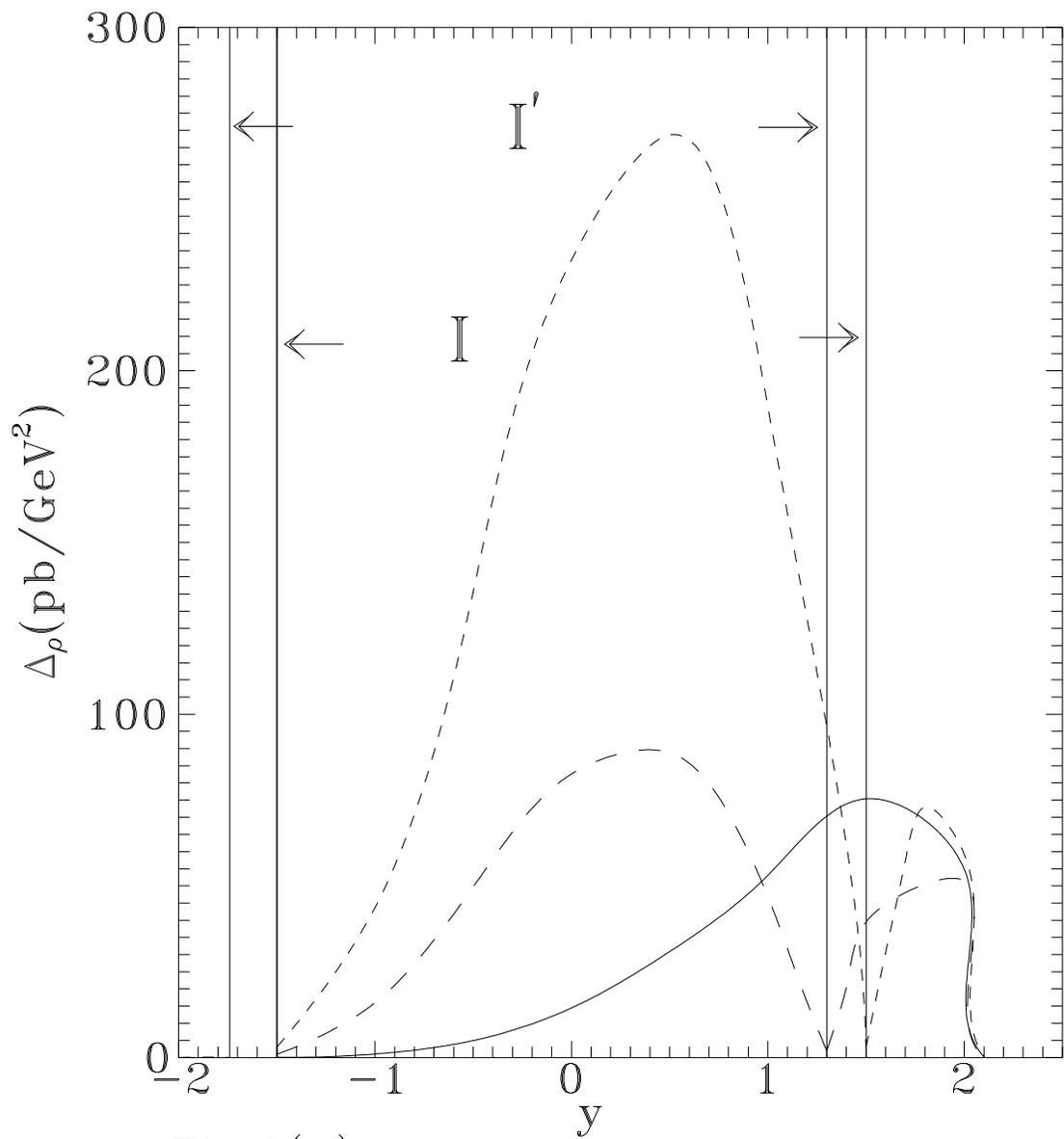
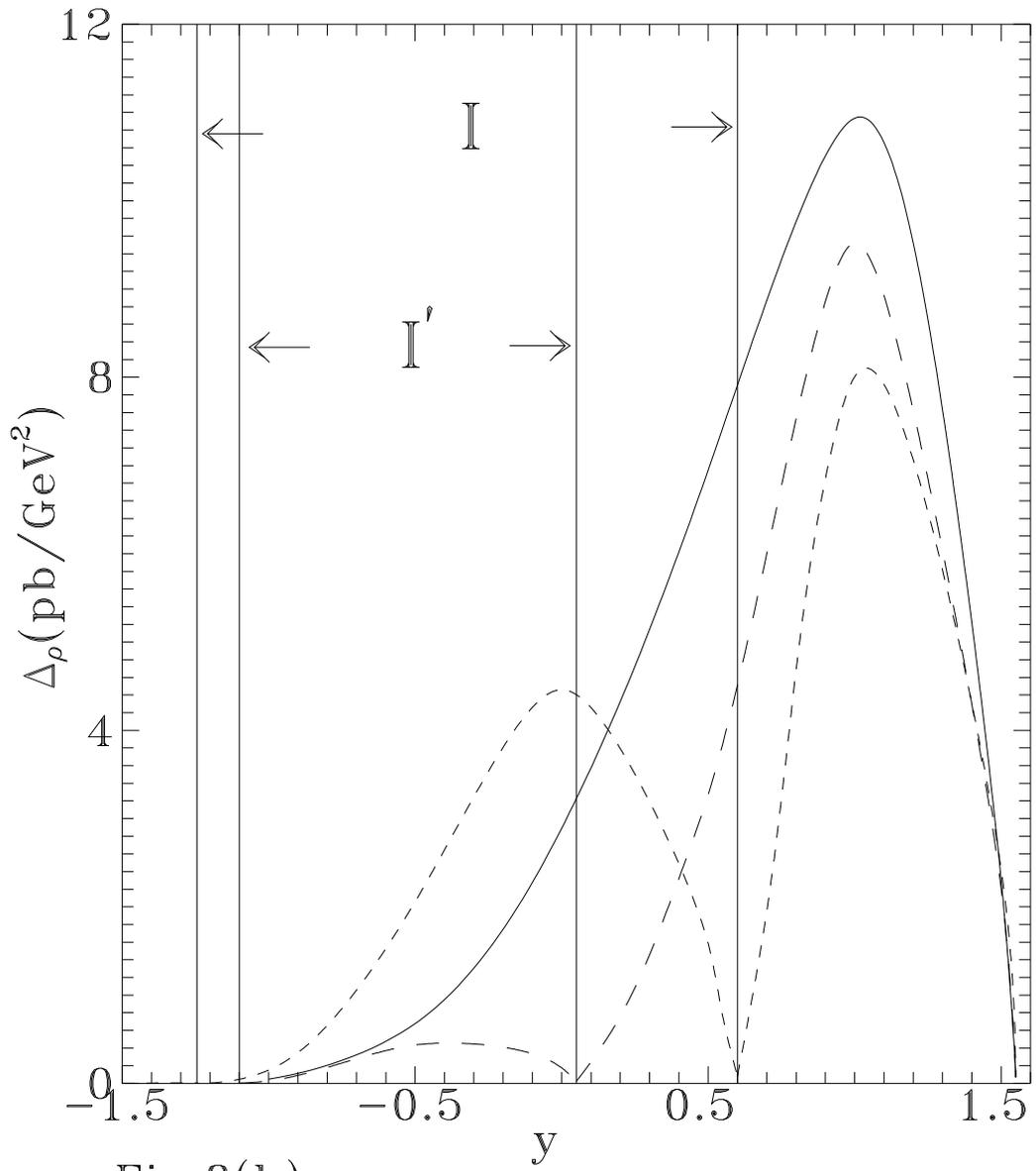
{\bf Fig.8} $\Delta_{\rho}$ dependence on the rapidity at $\sqrt{s}=25~GeV, 
p_{T}=3~GeV/c$ for a); at $\sqrt{s}=25~GeV, p_{T}=5~GeV/c$ for b). The solid 
curve corresponds to $\Delta_{\rho}^{LT}$, the long-dashed and short-dashed curves describe 
$\Delta_{\rho}^{tot}$ obtained using CZ and BB w.f., respectively. In regions
$I (BB~ w.f.)$ and $I' (CZ~ w.f.)$ the cross section difference 
$\Sigma_{\rho^{-}}^{tot}-\Sigma_{\rho^{+}}^{tot}$ is shown.
\end{document}